%% file: ICSME25.tex
\documentclass[conference]{IEEEtran}

\input{packages_macros}

\usepackage[numbers,sort&compress]{natbib}

\usepackage{listings}
\usepackage{caption}
\usepackage{color}

\usepackage{booktabs}
\usepackage{xcolor}
\usepackage{graphicx}
\usepackage[T1]{fontenc}

\usepackage{multirow}

\definecolor{custom-blue}{rgb}{0,0,1}

\begin{document}
% \pagestyle{plain}
% \pagenumbering{arabic}
\author{
    \IEEEauthorblockN{Abhiram Bellur}
    \IEEEauthorblockA{
    \textit{University of Colorado}\\
    % USA\\
    abhiram.bellur@colorado.edu}
    \and
    \IEEEauthorblockN{Fraol Batole}
    \IEEEauthorblockA{
    \textit{Tulane University}\\
    % USA\\
    fbatole@tulane.edu}
    \and
    \IEEEauthorblockN{Mohammed Raihan Ullah}
    \IEEEauthorblockA{
    \textit{University of Colorado}\\
    % USA\\
    raihan.ullah@colorado.edu}
    \and
    \IEEEauthorblockN{Malinda Dilhara}
    \IEEEauthorblockA{
    \textit{Amazon Web Services}\\
    % USA\\
    malwala@amazon.com}
    \and
    \IEEEauthorblockN{Yaroslav Zharov}
    \IEEEauthorblockA{
    \textit{JetBrains Research}\\
    % Germany\\
    yaroslav.zharov@jetbrains.com}
    \and
    \IEEEauthorblockN{Timofey Bryksin}
    \IEEEauthorblockA{
    \textit{JetBrains Research}\\
    % Cyprus\\
    timofey.bryksin@jetbrains.com}
    \and
    \IEEEauthorblockN{Kai Ishikawa}
    \IEEEauthorblockA{
    \textit{NEC Corporation}\\
    % Japan\\
    k-ishikawa@nec.com}
    \and
    \IEEEauthorblockN{Haifeng Chen}
    \IEEEauthorblockA{
    \textit{NEC Laboratories America}\\
    % USA\\
    haifeng@nec-labs.com}
    \and
    \IEEEauthorblockN{Masaharu Morimoto}
    \IEEEauthorblockA{
    \textit{NEC Corporation}\\
    % Japan\\
    m-morimoto@nec.com}
    \and
    \IEEEauthorblockN{Shota Motoura}
    \IEEEauthorblockA{
    \textit{NEC Corporation}\\
    % Japan\\
    motoura@nec.com}
    \and
    \IEEEauthorblockN{Takeo Hosomi}
    \IEEEauthorblockA{
    \textit{NEC Corporation}\\
    % Japan\\
    takeo.hosomi@nec.com}
    \and
    \IEEEauthorblockN{Tien N. Nguyen}
    \IEEEauthorblockA{
    \textit{University of Texas at Dallas}\\
    % USA\\
    tien.n.nguyen@utdallas.edu}
    \and
    \IEEEauthorblockN{Hridesh Rajan}
    \IEEEauthorblockA{
    \textit{Tulane University}\\
    % USA\\
    hrajan@tulane.edu}
    \and
    \IEEEauthorblockN{Nikolaos Tsantalis}
    \IEEEauthorblockA{
    \textit{Concordia University}\\
    % Canada\\
    nikolaos.tsantalis@concordia.ca}
    \and
    \IEEEauthorblockN{Danny Dig}
    \IEEEauthorblockA{
    \textit{University of Colorado, JetBrains Research}\\
    % USA\\
    danny.dig@colorado.edu}
}

\title{Together We Are Better: LLM, IDE and Semantic Embedding to Assist Move Method Refactoring}

\maketitle

\input{Files/abstract}

\input{Files/introduction}

\input{Files/motivating}

\input{Files/approach}

\input{Files/eval}
\input{Files/results}
\input{Files/threats}
\input{Files/related}
\input{Files/conclusion}

% \newpage

% \bibliographystyle{ACM-Reference-Format}
% \bibliography{references}

% \clearpage
\balance

\bibliographystyle{IEEEtran}
\bibliography{references}

\end{document}

%% file: packages_macros.tex
\usepackage{cite}
\usepackage{algorithmic}
\usepackage{graphicx}
\usepackage{textcomp}
\usepackage{xcolor}
\usepackage{amsmath,amsfonts}
\usepackage{algorithmic}
\usepackage{algorithm}
\usepackage{array}
\usepackage[caption=false,font=normalsize,labelfont=sf,textfont=sf]{subfig}
\usepackage{textcomp}
\usepackage{stfloats}
\usepackage{url}
\usepackage{verbatim}
\usepackage{graphicx}
\usepackage{subcaption}
\usepackage{xspace}
\usepackage{hyperref}
\usepackage{listings}
\usepackage{makecell}
\usepackage{booktabs}
\usepackage{algorithm}
\usepackage{caption}
\usepackage{threeparttable}
\usepackage{longtable}
\usepackage{tabularray}
\usepackage{cleveref}
\hypersetup{hidelinks}
\usepackage{color, colortbl}
\usepackage{balance}
\usepackage[inline]{enumitem}
\usepackage{calligra}
\usepackage{amsmath,amsthm}
\usepackage{tikz}
\usetikzlibrary{shapes.geometric, arrows}
\usetikzlibrary{positioning}
\usepackage{pifont}
\usepackage{tcolorbox}
\usepackage{makecell}
\usepackage{fancybox}
\usepackage{tcolorbox}
\usepackage{graphicx}
\usepackage{caption}
\usepackage{subcaption}

\usepackage{datetime2}

\newcommand{\tool}{\textsc{MM-assist}\xspace}

\newcommand{\code}[1]{{\texttt{\footnotesize{#1}}\xspace}}

\newcommand{\mm}{\textsc{MoveMethod}\xspace}

\theoremstyle{definition}
\newtheorem{definition}{Definition}[section]

\newcommand*\circled[1]{\tikz[baseline=(char.base)]{
            \node[shape=circle,draw,inner sep=1pt] (char) {#1};}}

\newcommand{\hallucinationRate}{80\%\xspace}
\newcommand{\hallucinationRateSynthetic}{{80\%}\xspace}

\newcommand{\feTruth}{\textsc{feTruth}\xspace}
\newcommand{\JMove}{\textsc{JMove}\xspace}
\newcommand{\HMove}{\textsc{HMove}\xspace}

\newcommand{\totalRealWorldModern}{210\xspace} % = totalRealWorldModernInstance + totalRealWorldModernStatic
\newcommand{\totalRealWorldModernInstance}{108\xspace} % instance method count
\newcommand{\totalRealWorldModernStatic}{102\xspace} % static method count

\newcommand{\totalSyntheticAndRealRecommendations}{2016\xspace} % total recommendations on Synthetic+Real recommendations (from Fig 3)

\newcommand{\llmAvgNumSuggestions}{6\xspace} 
\newcommand{\llmFinalUsefullPercentageExtendedCorpus}{20\%\xspace} % Abhiram to compute
\newcommand{\totalmmAssistSuggestions}{1293\xspace} % total number of suggestion by mm-assist
\newcommand{\typeOneHallucination}{431\xspace} % total number of typ-1 halls.
\newcommand{\typeTwoHallucination}{275\xspace} % total number of suggestion by mm-assist
\newcommand{\typeThreeHallucination}{320\xspace} % total number of suggestion by mm-assist

\newcommand{\RecallAtOneInstance}{{67\%}\xspace} % From Table 2: Evaluation on Synthetic corpus
\newcommand{\RecallAtThreeInstance}{{75\%}\xspace} % From Table 2: Evaluation on Synthetic corpus
\newcommand{\rwisRecallMOneVanilla}{{52\%}\xspace}
\newcommand{\rwisRecallMTwoVanilla}{{71\%}\xspace}
\newcommand{\rwisRecallMThreeVanilla}{{83\%}\xspace}
\newcommand{\rwisRecallCOneVanilla}{{67\%}\xspace}
\newcommand{\rwisRecallCTwoVanilla}{{67\%}\xspace}
\newcommand{\rwisRecallCThreeVanilla}{{67\%}\xspace}
\newcommand{\rwisRecallMCTwoVanilla}{{54\%}\xspace}
\newcommand{\rwisRecallMCThreeVanilla}{{58\%}\xspace}

\newcommand{\rwisRecallMOne}{{75\%}\xspace}
\newcommand{\rwisRecallMTwo}{{92\%}\xspace}
\newcommand{\rwisRecallMThree}{{95\%}\xspace}
\newcommand{\rwisRecallCOne}{{85\%}\xspace}
\newcommand{\rwisRecallCTwo}{{89\%}\xspace}
\newcommand{\rwisRecallCThree}{{89\%}\xspace}
\newcommand{\rwisRecallMCOne}{{68\%}\xspace}
\newcommand{\rwisRecallMCTwo}{{80\%}\xspace}
\newcommand{\rwisRecallMCThree}{{80\%}\xspace}

\newcommand{\rwilRecallMOneVanilla}{{15\%}\xspace}
\newcommand{\rwilRecallMTwoVanilla}{{23\%}\xspace}
\newcommand{\rwilRecallMThreeVanilla}{{24\%}\xspace}
\newcommand{\rwilRecallCOneVanilla}{{44\%}\xspace}
\newcommand{\rwilRecallCTwoVanilla}{{44\%}\xspace}
\newcommand{\rwilRecallCThreeVanilla}{{44\%}\xspace}
\newcommand{\rwilRecallMCTwoVanilla}{{8\%}\xspace}
\newcommand{\rwilRecallMCThreeVanilla}{{8\%}\xspace}

\newcommand{\rwilRecallMOne}{{36\%}\xspace}
\newcommand{\rwilRecallMTwo}{{38\%}\xspace}
\newcommand{\rwilRecallMThree}{{47\%}\xspace}
\newcommand{\rwilRecallCOne}{{75\%}\xspace}
\newcommand{\rwilRecallCTwo}{{80\%}\xspace}
\newcommand{\rwilRecallCThree}{{80}\xspace}
\newcommand{\rwilRecallMCOne}{{30\%}\xspace}
\newcommand{\rwilRecallMCTwo}{{31\%}\xspace}
\newcommand{\rwilRecallMCThree}{{36\%}\xspace}

\newcommand{\HMOVErwisRecallMOne}{23\%\xspace}
\newcommand{\HMOVErwisRecallMTwo}{37\%\xspace}
\newcommand{\HMOVErwisRecallMThree}{40\%\xspace}
\newcommand{\HMOVErwisRecallCOne}{37\%\xspace}
\newcommand{\HMOVErwisRecallCTwo}{43\%\xspace}
\newcommand{\HMOVErwisRecallCThree}{47\%\xspace}
\newcommand{\HMOVErwisRecallMCOne}{17\%\xspace}
\newcommand{\HMOVErwisRecallMCTwo}{30\%\xspace}
\newcommand{\HMOVErwisRecallMCThree}{33\%\xspace}

\newcommand{\HMOVErwilRecallMOne}{9\%\xspace}
\newcommand{\HMOVErwilRecallMTwo}{15\%\xspace}
\newcommand{\HMOVErwilRecallMThree}{20\%\xspace}
\newcommand{\HMOVErwilRecallCOne}{15\%\xspace}
\newcommand{\HMOVErwilRecallCTwo}{20\%\xspace}
\newcommand{\HMOVErwilRecallCThree}{24\%\xspace}
\newcommand{\HMOVErwilRecallMCOne}{4\%\xspace}
\newcommand{\HMOVErwilRecallMCTwo}{5\%\xspace}
\newcommand{\HMOVErwilRecallMCThree}{5\%\xspace}

\newcommand{\mmAssistRuntimeAvg}{{27.5}\xspace} % average runtime of mm-assist

\newcommand{\llmRuntimeAvg}{{9}\xspace} % average runtime to get response from llm

\newcommand{\numUserParticipants}{30\xspace} % ~43 enrolled participants, only 30 sent data

\newcommand{\percentPositivePerScreen}{82.8\%\xspace}
\newcommand{\totalClassesUser}{350\xspace}
\newcommand{\totalPositiveClassesUser}{290\xspace}

\newcommand{\totalAppliedUser}{216\xspace}
\newcommand{\avgAppliedUser}{7\xspace}

\newcommand{\percentPositiveRatingSurvey}{{80\%}\xspace} % TODO: update if more respond: currently its 10/12

\newcommand{\overwhelmRecoCount}{{57}\xspace}

%% file: Files/abstract.tex
\begin{abstract}

\mm is a hallmark refactoring. 
%to remedy the lack of code modularity and remove several code smells that contribute to technical debt.
%While all the leading IDEs provide support for executing this refactoring automatically, they do not recommend which methods should be moved and where. 
Despite a plethora of research tools that recommend which methods to move and where, % by optimizing software quality metrics, 
these recommendations do not align with how expert developers perform \mm. 
%However, static analysis-based approaches require experts to specify thresholds, whereas ML-based approaches require training or constant re-training as coding standards and practices evolve. 
%However, previous research recommends refactorings that do not align with how expert developers perform such refactorings in practice. 
Given the extensive training of Large Language Models and their reliance upon \emph{naturalness of code}, they should expertly recommend which methods are \emph{misplaced} in a given class and which classes are better hosts. %for such misplaced methods. 
%Moreover, their recommendations should align with experts.
{Our formative study of \totalSyntheticAndRealRecommendations LLM recommendations revealed that LLMs give expert suggestions, yet they are unreliable: up to \hallucinationRateSynthetic of the suggestions are hallucinations.}

We introduce the first LLM fully powered assistant for \mm refactoring that automates its whole end-to-end lifecycle, from recommendation to execution. We designed novel solutions that automatically filter LLM hallucinations using static analysis from IDEs and a novel workflow that requires LLMs to be \emph{self-consistent, critique, and rank} refactoring suggestions. 
%Moreover, \mm refactoring requires global, project-level reasoning to determine the best target classes where to relocate a misplaced method. 
As \mm refactoring requires global, project-level reasoning, we solved the limited context size of LLMs by employing \emph{refactoring-aware retrieval augment generation} (RAG). 
Our approach, \tool,  synergistically combines the strengths of the LLM, IDE, static analysis, and semantic relevance. 
%\tool generates candidates, filters LLM hallucinations, validates and ranks recommendations, and then finally executes the correct refactoring based on user approval. 
In our thorough,  multi-methodology empirical evaluation, we compare \tool with the previous state-of-the-art approaches. \tool significantly outperforms them: (i) 
on a benchmark widely used by other researchers, our Recall@1 and Recall@3 show a {1.7x} improvement;
(ii) on a corpus of \totalRealWorldModern recent refactorings from Open-source software, our Recall rates improve by {at least 2.4x}. 
%our Recall@1 and Recall@3 are {\RecallAtOneInstance and \RecallAtThreeInstance}, respectively, which is a 2x improvement over the state-of-the-art approaches (33\% and 37\%). 
%Moreover, we~extend the corpus used by previous researchers with \totalRealWorldModern actual refactorings performed by Open-source software developers in 2024; \tool achieves even more significant improvements over previous state-of-the-art tools, our Recall@1 is \rwisRecallMCOne, and Recall@3 is \rwisRecallMCThree, compared to \feTruthProcessedRealWorld\% and \HMOVErwisRecallMCThree, i.e., an almost 3.5x improvement. 
Lastly, we conducted a user study with \numUserParticipants experienced participants who used \tool to refactor their own code for one week. They rated \percentPositivePerScreen of \tool recommendations positively. 
This shows that \tool is both effective and useful.

\end{abstract}

% \begingroup
% \renewcommand\thefootnote{}
% \footnotetext{* next to the authors indicates that they contributed equally to this work.}
% \endgroup

%% file: Files/introduction.tex
\section{Introduction}
\label{sec:intro}

\mm is a key refactoring~\citep{fowler1997refactoring} that~relocates a misplaced method to a more suitable class. A method is \emph{misplaced} if it interacts more with another class' state than its own. 
\mm improves modularity by aligning methods with relevant data, enhances cohesion, and reduces coupling. It removes code smells like FeatureEnvy~\citep{Tsantalis:2009:IMM:1591905.1592367},~GodClass~\citep{BAVOTA2011397}, DuplicatedCode~\citep{10.1145/2950290.2950317}, and MessageChain~\citep{fowler1997refactoring}, reducing technical debt. It ranks among the top-5 most common refactorings~\citep{negara2013,murphy2012,refactoringminer2}, both manually and automatically performed.

%\mm is a key refactoring technique~\citep{fowler1997refactoring} that moves a misplaced method from its original class (source) to a more suitable target class. A method is considered \emph{misplaced} when it doesn't interact with its own class's state or relies more on the state of another class. Developers use \mm to improve modularity by grouping related behavior (methods) with the data (fields) it affects, which enhances class cohesion and reduces coupling between classes. \mm helps eliminate several code smells, such as FeatureEnvy~\citep{Tsantalis:2009:IMM:1591905.1592367}
%, GodClass~\citep{BAVOTA2011397}%
%, DuplicatedCode~\citep{10.1145/2950290.2950317}
%, and MessageChain~\citep{fowler1997refactoring}. Thus, \mm reduces technical debt and is one of the top-5 most common refactorings~\citep{negara2013,murphy2012,refactoringminer2}, in both manual and automated settings.

% it consists of 4 phases: identification, parameterization, and execution
% The IDEs automate the execution, the researchers address recommendation.

The \mm lifecycle has four phases: (i) identifying a misplaced method \code{m} in its host class \code{H}, (ii) finding a suitable target class \code{T}, (iii) ensuring refactoring pre- and post-conditions to preserve behavior, and (iv) executing the transformation. Each phase is challenging—identifying candidates requires understanding design principles and the codebase, while checking preconditions~\citep{OpdykeThesis, Tsantalis:2009:IMM:1591905.1592367} demands complex static analysis. The mechanics involve relocating \code{m}, updating call sites, and adjusting accesses. Due to these complexities, existing solutions are incomplete; IDEs handle preconditions and mechanics, while research tools aim to identify opportunities.

The research community has proposed various approaches~\citep{Tsantalis:2009:IMM:1591905.1592367, Bavota:2014:MethodBook, TERRA201819JMove, 10.1145/3242163.3242171, 10.1145/3387940.3392191, RMove:2022, Liu2018CodeSmells, 8807230, feTruth} for identifying misplaced methods or recommending new target classes, typically optimizing software quality metrics like cohesion and coupling. These approaches fall into three categories: (i) static analysis-based~\citep{Tsantalis:2009:IMM:1591905.1592367, Bavota:2014:MethodBook, TERRA201819JMove}, (ii) machine learning classifiers~\citep{10.1145/3242163.3242171, 10.1145/3387940.3392191, RMove:2022}, and (iii) deep learning-based~\citep{Liu2018CodeSmells, 8807230, feTruth}. However, static analysis relies on expert-defined thresholds, and ML/DL methods require continual retraining as coding standards evolve, often diverging from real-world development practices.

% \shorten{However, static analysis approaches rely on expert-defined thresholds, and ML/DL methods need constant retraining as coding standards and practices evolve. Additionally, their recommendations often don't align with how expert developers perform \mm in practice.}

We hypothesize that achieving good software design that is easy to understand and resilient to future changes is a \emph{balancing act between science} (e.g., metrics, design principles) and \emph{art} (e.g., experience, expertise, and intuition about what constitute good abstractions). This can explain why refactorings that optimize software quality metrics are not always accepted in practice~\cite{MindTheGap, look_ahead_rohit,automated_variable_bavota,why_do_bavota,why_we_marco,software2,mlcodechanges,pyevolve,fakhoury2019improving,scalabrino2019automatically}.

% Our approach: use LLMs to find methods out of scope. Why it makes sense, what are some challenges (lack of global context --limited size for context, hallucinations).
%In this paper, we present the first approach to automate the full lifecycle of \mm refactoring powered by Large Language Models (LLMs).  We hypothesize that given the huge pre-training of LLMs on billions of methods and their reliance upon the \emph{naturalness of code}, they should be better at recommending which methods are \emph{misplaced} in a given class and which classes are better hosts for such misplaced methods. We hypothesize that LLM  recommendations better align with experts. In our formative study of LLM recommendations, we discovered the LLMs are quite prolific in coming up with recommendations. On average, they produce \todo{XYZ recommendations} per class. However, there are two major challenges that we need to overcome to make such an LLM-based solution practical. 

In this paper, we introduce the first approach to automate the entire \mm refactoring lifecycle using Large Language Models (LLMs). We hypothesize that, due to their extensive pre-training on billions of methods and their reliance on the \emph{naturalness of code}, 
% LLMs should excel at identifying \emph{misplaced} methods and suggesting better host classes. 
LLMs can generate an abundance of \mm recommendations.
We also expect LLM recommendations to better align with expert practices. 
% We use GPT-4o, which has demonstrated superior performance on programming tasks~\citep{chiang2024chatbot,arenaLeaderBoard} and is widely adopted in developer productivity tools~\citep{Cursor,copilot}. 
In our formative study, we found LLMs ({GPT-4o, in particular}) are prolific in generating suggestions, averaging \llmAvgNumSuggestions recommendations per class. However, two major challenges must be addressed to make this approach practical.

% First challenge: hallucinations
%First, LLMs are known to produce \emph{hallucinations}, i.e., recommendations that seem plausible but are deeply flawed. In our formative study of \totalSyntheticAndRealRecommendations LLM recommendations, we found three kinds %\todo{four kinds} of LLM hallucinations: (i) non-existent target classes (i.e., the LLM-suggested target class does not exist in the current project), (ii) unfeasible target classes (i.e., refactoring is unfeasible as it is impossible to access the Target class from the Host class), (iii) wrong method picked as misplaced in a Host class. In our formative study, we discovered that \hallucinationRate to \hallucinationRateSynthetic of LLM recommendations (for static and instance methods, respectively) are hallucinations. Thus, vanilla LLM approaches need significant further processing to increase their reliability.

First, LLMs produce \emph{hallucinations}, i.e., recommendations that seem plausible but are flawed. In our formative study of \totalSyntheticAndRealRecommendations LLM recommendations, {we identified three types of hallucinations ({e.g., recommendations where the target class does not exist}), and found that up to \hallucinationRate of LLM recommendations are hallucinations.}
% we identified three types of hallucinations: (i) LLM suggests target classes that do not exist in the project, (ii) it is impossible to move a method in the suggested target class, 
% since the Target class cannot be accessed from the Host class
% and (iii) LLM identifies invalid methods as misplaced in a host class. {Our findings reveal that up to \hallucinationRate} 
% to \hallucinationRateSynthetic 
% of LLM recommendations 
% (for static and instance methods, respectively) 
% are hallucinations. This requires further processing to enhance LLM reliability. 
% How we solve these challenges: The Hallucinations
 We discovered novel ways to automatically eliminate LLM hallucinations, by complementing LLM reasoning (i.e., {\em the~creative, non-deterministic, and artistic part} akin to human naturalness) with static analysis embedded in the IDE (i.e., {\em the rigorous, deterministic, scientific part}). 
 {We utilized code-trained vector embeddings from AI models to identify misplaced methods, and used refactoring preconditions~\citep{OpdykeThesis} in existing IDEs (IntelliJ IDEA) to effectively remove \emph{all} LLM hallucinations.}
 % Mature refactoring implementations in existing IDEs like IntelliJ IDEA contain a plethora of static analysis checks (called refactoring preconditions~\citep{OpdykeThesis}) that must be true before a method can be safely moved to another~class. We use these refactoring preconditions to validate LLM recommendations. %, thus helping remove \todo{\typeThreeHallucinationRate} of the hallucinations. 
% Moreover, before asking the LLM for recommendations, we designed a filtering stage by using code embedding to identify the least compatible \code{m} in \code{H}.
% Moreover, we compute semantic similarity between a method and its host class using code-trained embeddings, identifying the least cohesive methods and focus the LLM on these methods.
% These stages effectively removed \emph{all} LLM hallucinations.
We present these techniques in Section~\ref{subsection:moveMethod}.

% Describing the problem and solution for challenge 2: limited size context
% Para 1: How we solve these challenges: RAG approach
Second, \mm refactoring requires global, project-level reasoning to determine the best target classes where to relocate a misplaced method. However, passing an entire project in the prompt is beyond the limited context window of current LLMs~\citep{FocusedTransformer:NIPS23}. 
{Even with larger window sizes}, passing the whole project as context introduces noise and redundancy, as not all classes are relevant; instead this further distracts the LLM~\citep{FocusedTransformer:NIPS23,NeedleInHaystack}.
% We overcame the limited context size of LLMs by using \emph{retrieval augmented generation} (RAG). ... \todo{Fraol to add a sentence or two about using the RAG approach for handling limited context size.}
{We address the limited context window of LLMs by using \emph{retrieval augmented generation} (RAG) to enhance the LLM's input.}
% with relevant project-specific information for better decision-making. 
% However, a naive approach of retrieving all similar classes would be ineffective and worsen the hallucination problem. 
% Instead \commentb{of blindly retrieving classes}, we first apply IDE-based static analysis to identify mechanically feasible target classes, significantly narrowing the search space and reducing hallucinations. While static analysis ensures feasibility, we also leverage semantic relevance to find a suitable target class. We utilize VoyageAI~\citep{voyageAI}, which has demonstrated state-of-the-art performance in code-related tasks~\citep{wang2024coderag}.
% For this purpose, we use VoyageAI~\citep{voyageAI}, an embedding trained on code that integrates natural language, programming language, and data-flow information for nuanced semantic understanding. 
Our two-step retrieval process combines mechanical feasibility (IDE-based static analysis) with semantic relevance (VoyageAI~\citep{voyageAI}), enabling our approach to make informed decisions and perform global project-level reasoning. We coin this approach \emph{refactoring-aware retrieval augmented generation}, which addresses LLM hallucinations and context limitations while fulfilling the specific needs of \mm refactoring (see Section~\ref{subsection:targetClass}).

%We overcame the limited window context size of LLMs by using \emph{retrieval augmented generation} (RAG). RAG is used to augment the LLM's input with relevant project-specific information for more informed decision-making. However, a naive approach of retrieving all similar classes would be ineffective and exacerbate the hallucination problem. Instead, we first employ IDE-based static analysis to identify mechanically feasible target classes, significantly reducing the search space and mitigating hallucinations. While static analysis ensures mechanical feasibility, it is also essential to consider semantic relevancy to find a suitable target class. To do so, we use VoyageAI~\citep{voyageAI}, an embedding specifically trained on code that combines natural language, programming language, and data-flow information for nuanced semantic understanding. This two-step process balances mechanical feasibility with semantic relevance, enabling \tool to make informed decisions and perform global, project-level reasoning. We coin this approach \emph{refactoring-aware retrieval augmented generation}, as it addresses both the challenges of LLM hallucinations and context limitations while meeting the specific requirements of \mm refactoring.

% Implementation: plugin

We designed, implemented, and evaluated these novel solutions as an IntelliJ IDEA plugin for Java, \tool (Move Method Assist). It synergistically combines the strengths of the LLM, IDE, static analysis, and semantic relevance. \tool generates candidates, filters LLM hallucinations, validates and ranks recommendations, and finally executes the correct refactoring based on user approval using the IDE.

%Evaluation: part 1
We designed a comprehensive, multi-methodology evaluation of \tool to corroborate, complement, and expand research findings: formative study, comparative study, replication of real-world refactorings, repository mining, user/case study, and questionnaire surveys. 
% \shorten{Among others, our formative study of \totalSyntheticAndRealRecommendations LLM recommendations reveals the strengths and weaknesses of using LLMs for recommending \mm refactorings. To quantify the improvements of \tool over the vanilla LLM solution, we conducted an ablation study that shows \tool brings significant improvements. 
% Moreover,} 
{We} compare \tool with the previous best in class approaches in their respective categories: 
\JMove~\citep{TERRA201819JMove} -- uses static analysis, \feTruth~\citep{feTruth} -- uses Deep Learning, and \HMove~\citep{cui2024three} -- uses graph neural network to suggest moves and LLM to verify refactoring preconditions. These have been shown to outperform all previous \mm recommendation tools. Using a synthetic corpus widely used by previous approaches, we found that our tool significantly outperforms them: for class instance methods, 
our Recall@1 and Recall@3 are 
{\RecallAtOneInstance and \RecallAtThreeInstance}, respectively, {which is almost 1.75x improvement over previous state-of-the-art approaches (40\% and 42\%).}
Moreover, we extend the corpus used by previous researchers with \totalRealWorldModern actual refactorings that we mined from OSS repositories in 2024 (thus avoiding LLM data contamination), {containing both instance and static methods}. We compared against \JMove, \HMove and \feTruth on this real-world oracle, and found that \tool significantly outperforms them. Our 
% Recall@1 is \rwisRecallMCOne, and
{Recall@3 is \rwisRecallMCThree, compared to \HMOVErwisRecallMCThree for the best baseline, {\HMove, a 2.4x improvement.}
}% As \JMove does not support moving static methods (because the search space is too large), we compare with \feTruth. \tool achieves even more significant improvements over previous tools, our Recall@1 is \RecallAtOneStatic, and Recall@3 is \RecallAtThreeStatic, compared to \todo{BestFeTruth\%}. 
This shows that \tool's recommendations better align with developer best practices.

\input{Figures/motivatingExample}

% \shorten{To recommend which method(s) to move from a class, previous tools require analyzing an entire project -- which takes between 2 hours for medium projects to more than 10 hours for 1M LOC projects,  and they overwhelm the user with up to 40 recommendations per class. \tool is modular: it takes on average 30 seconds (even on projects with tens of thousands of classes), and it shows no more than 3 high quality recommendations per class. Thus, \tool is practical.}

{Whereas existing tools often require several hours to analyze a project and overwhelm developers with up to {
% 40
\overwhelmRecoCount recommendations} to analyze per class, \tool needs only about 30 seconds—even for tens of thousands of classes—and provides no more than 3 high-quality {recommendations} per class.}

% \shorten{To better understand how developers use \tool in practice, we recruited \numUserParticipants experienced participants who used it on their own code for a week and provided telemetry data. Unlike previous studies that had participants assess recommendations on unfamiliar third-party code, our study allows participants to evaluate recommendations on code they deeply understand. Results show that \percentPositivePerScreen of participants rated \tool's recommendations positively and preferred our LLM-based approach over classic IDE workflows.}
{In a study where \numUserParticipants experienced participants used \tool on their own code for a week, \percentPositivePerScreen rated its recommendations positively and preferred our LLM-based approach over classic IDE workflows.}
{One participant remarked, \emph{``I am fairly skeptical when it comes to AI in my workflow, but still excited at the opportunity to delegate grunt work to them.''}}

%To further shed light on how developers use \tool in practice, we recruited \numUserParticipants experienced participants who used \tool on their own authored code for one week and sent us their telemetry data. While previous studies asked participants to assess recommendations on some third-party open-source code they are not familiar with, our user study ensures that participants are most likely to accurately assess the quality of the recommendations on the code they deeply understand. The results show that \percentPositivePerScreen of participants gave \tool's recommendations a positive rating, and they prefer our LLM-based approach to the classic workflow in the IDEs. Among others, participants said \emph{``I am fairly skeptical when it comes to AI in my workflow, but still excited at the opportunity to delegate grunt work to them''}.   

% Contributions
This paper makes the following contributions:
\begin{itemize}
    \item{\textbf{Approach.}} We present the {first end-to-end LLM-powered assistant}
    % that supports the lifecycle of recommending and applying 
    for \mm. Our approach advances key aspects: (i)
    % it ensures the 
    recommendations are feasible and executed correctly,
    % upon user approval, 
    (ii) it requires no user-specified thresholds or model (re)-training, making it future-proof as LLMs evolve, and (iii) it handles both instance 
    % — like other solutions — 
    and static methods (avoided by others due to large search space).  
   
    \item{\textbf{Best Practices.}} We discovered a new set of best practices to overcome the LLM limitations when it comes to refactorings that require global reasoning. We automatically filter LLM hallucinations and conquer the LLM's limited context window size using refactoring-aware RAG.  
    \item{\textbf{Implementation.}} {We designed, implemented, and evaluated these ideas in an IntelliJ plugin, \tool, that works on Java code. It addresses practical considerations for tools used in the daily workflow of developers.}
    %It synergistically combines the strengths of the LLM, IDE, static analysis, and semantic relevance. 

    \item{\textbf{Evaluation.}} We thoroughly evaluated \tool, and it outperforms previous best-in-class approaches. We also created an oracle replicating actual refactorings done by OSS developers, where \tool showed even more improvements. {Our user study confirms that} 
    % developers prefer our LLM-based assistant, demonstrating that 
    \tool aligns with and replicates real-world expert logic.

    \item{\textbf{Replication.}} {We freely release \tool, the datasets we used in the experiments, LLM prompts, demo, etc, so others can build upon these~\citep{ReplicationPackage}}.
\end{itemize}

%% file: Figures/motivatingExample.tex
\begin{figure*}[t]
    \centering    
    \includegraphics[width=0.8\linewidth,trim=0cm 0cm 0cm 0cm]{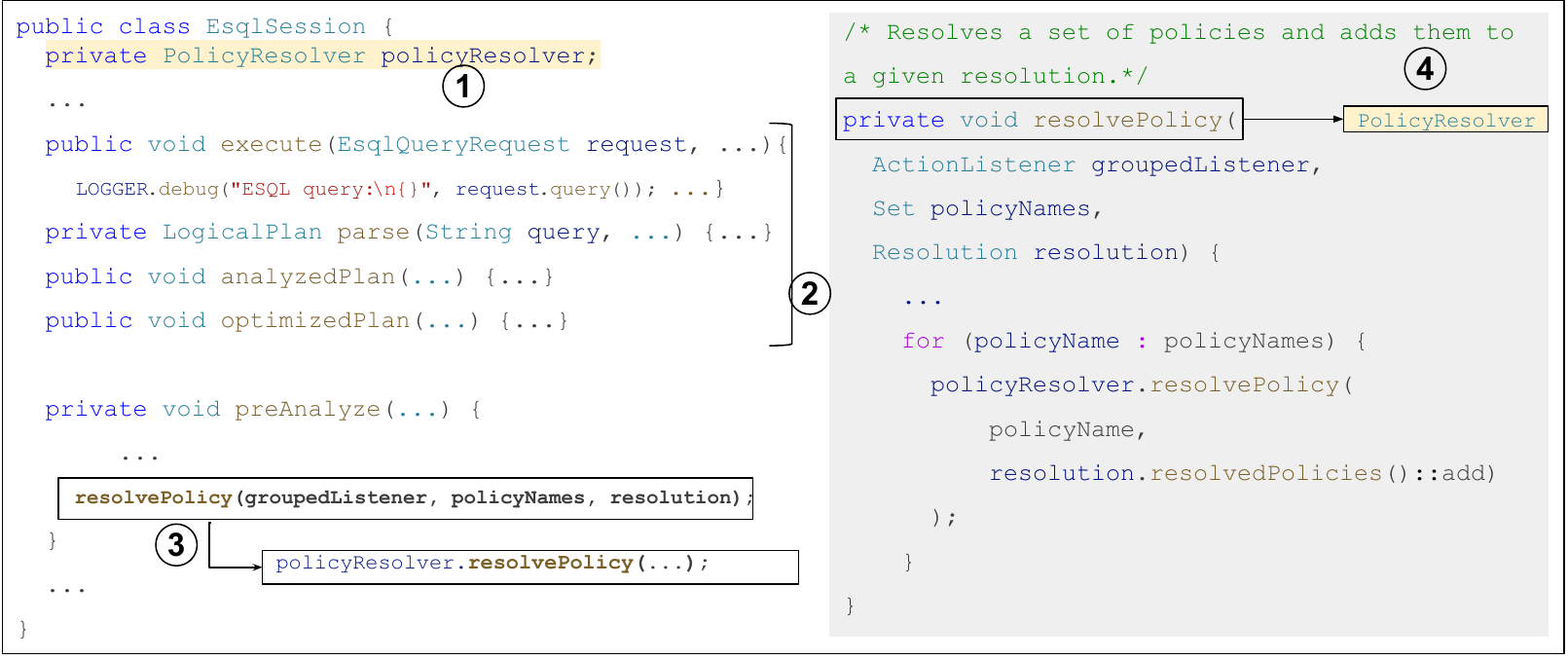}
    % \vspace{-15pt}
    \caption{A real-world example demonstrating a \mm on \code{resolvePolicy} performed by developers in the {\em Elasticsearch} project, commit 876e7015}
    \label{fig:motivatingExample}
\end{figure*}

%% file: Files/motivating.tex
\section{Motivating Example}
\label{sec:motivating}

% We illustrate the challenges of of recommending \mm refactoring using a real-world example (see~ \Cref{fig:motivatingExample}), from the Restlet Framework~\citep{mmcommit}. The \code{computeSectionName} method (See \circled{1} in \Cref{fig:motivatingExample}), originally located in the \code{SwaggerReader} class, transforms an API declaration path into a section name. This method's placement within the class  \code{SwaggerReader} may reduce the class modularity and cohesiveness as the method performs string manipulation, a functionality distinctly different from the rest of the methods within the class. The developers recognized the misplacement of this method as they refactored the code by moving \code{computeSectionName} into a dedicated utility class (\circled{4} in \Cref{fig:motivatingExample}), \code{SwaggerUtils}, thereby reducing the technical debt.

We illustrate the challenges of recommending \mm using a real-world refactoring that occurred in the \textit{Elasticsearch} project -- a distributed open-source search and analytics engine. We illustrate the refactoring in~\Cref{fig:motivatingExample}, and the full commit can be seen in ~\citep{mmcommit}. The \code{resolvePolicy} method (See \circled{4} in \Cref{fig:motivatingExample}), originally part of the \code{EsqlSession} class, is misplaced. While \code{EsqlSession} handles parsing and executing queries, \code{resolvePolicy} is responsible for resolving and updating policies. Specifically, \code{resolvePolicy} accesses the field \code{policyResolver} (See \circled{1}) and parameters like \code{groupedListener}, \code{policyNames}, and \code{resolution}. Recognizing this misalignment, the developers refactored the code by moving \code{resolvePolicy} to the \code{PolicyResolver} class (not shown in the figure due to space constraints), updating the method body accordingly, and modifying the call sites (See \circled{3}). After the refactoring, both \code{EsqlSession} and \code{PolicyResolver} became more cohesive.

% Automating the identification of such refactoring opportunities is important but presents significant challenges. To that end, we applied a state-of-the-art technique to detect feature envy called feTruth~\citep{feTruth} to this scenario, but they failed to recommend the move-method refactoring performed by the developers. These techniques, which typically rely on metrics such as coupling and cohesion, did not capture the semantic mismatch between \code{computeSectionName} and its original class. 

Automatically identifying such refactoring opportunities is essential for maintaining software quality, but it poses significant challenges for existing tools. We first applied \HMove~\citep{cui2024three}, the state-of-the-art \mm technique. 
\HMove, a classification tool, takes in $<$method, targetClass$>$ pairs, and gives a probability score indicating whether to move the method. After computing all 158 pairs of inputs, \HMove executed for 1.5 hours and 
generated 36 \mm recommendations.
% \HMove, after computing all 158 pairs of inputs. It took 1.5 hours to run and made 36 \mm recommendations. 
Its first and second highest recommendations to move are \code{subfields} and \code{execute}.
% After \commentD{running it on 158 different pairs} of (method, target class) for 1.5 hours, \todo{describe the scenario}.
% Unfortunately, due to internal bugs (failing to deal with new Java language features), \HMove did not run successfully in this case.} 
% Next, we applied \feTruth~\citep{feTruth} \todo{the previous state-of-the-art} Feature Envy detection technique, \feTruth~\citep{feTruth}}, 
% to this scenario. 
Its third recommendation to move is \code{resolvePolicy}.
However, \HMove recommended moving \code{resolvePolicy} to the class \code{FunctionRegistry}, which is not an appropriate fit, as the method {does not interact} with it. {This illustrates major shortcomings of classification-based tools like \HMove: they need to be triggered on {lots (avg. 145)} of $<$method, targetClass$>$ pairs, which means long runtime. Furthermore, they overwhelm the users with so many recommendations to analyze, testing the developer's patience and endurance. Moreover, they don't provide actionable steps. 
% not only do they require a large amount of time and resources to execute, but also provide an overwhelming amount of recommendations, testing both developers' patience and \todo{endurance}. Furthermore, they fail to provide actionable recommendations.
% Instead, it suggested moving {\code{execute(...)}}, which is a method highly cohesive with the class's primary responsibilities.
% which \todo{is a core functionality} of the class.
% and a suggestion that does not match the developers' actual refactoring. 
% {This shows that DL models trained on real-world datasets (such as those used by \HMove) may not generalize well in cases where domain-specific knowledge or project-specific design play key roles.}
}
% \shorten{Since \feTruth relies on training DL models for refactoring using real-world datasets, this result highlights that 
% even specialized DL 
% such models may not
% capture all refactoring scenarios 
% generalize well in cases
% where domain-specific knowledge or project-specific design patterns play key roles.}

Next, we ran \JMove~\citep{TERRA201819JMove}, a state-of-the-art \mm recommendation tool that solely relies on static analysis.
% to compute program dependencies. 
To analyze the whole project \JMove requires {12+ hours}. To speed up \JMove, we ran it on a sub-project of Elasticsearch containing \code{EsqlSession}. 
{After {it finished running for 30 minutes} on a sub-project of Elasticsearch,
% Unfortunately, 
\JMove did not produce any recommendations for the \code{EsqlSession} class}.
% Moreover, \JMove took {30} minutes to report any results as it needed to analyze the entire sub-project and create program dependencies. 
{This highlights a major shortcoming of existing static-analysis based tools like \JMove: they need to analyze the entire project-structure and compute program dependencies -- thus they do not scale to
% as developers would run out of patience when executing tools  
medium to large-size projects like \code{Elasticsearch} (800K LOC).}

%\todo{@Fraol/Abhiram: please verify this sentence, please run JMove to see whether it produces anything}.

% This limitation potentially stems from feTruth's training on datasets collected using RefactoringMiner, which is known to produce false positives in certain scenarios. For instance, RefactoringMiner may incorrectly identify class deletions followed by recreations in different locations as move method refactorings. Despite the authors' efforts to clean the dataset, these inherent biases in the mining process might have influenced the model's learning, leading to overdetection of move method opportunities in classes like \code{SwaggerReader}. This case highlights the challenges in training machine learning models for refactoring tasks using automatically mined datasets, even with careful data cleaning processes.

Next, we explored the potential of Large Language Models (LLMs) to recommend \mm refactoring. We used GPT-4o, a state-of-the-art LLM developed by OpenAI~\citep{ChatGPT}, and prompted it with the content of the \code{EsqlSession} class, asking:  
{\em ``Identify methods that should move out of the \code{EsqlSession} class and where?''}. Our result highlighted both the strengths and limitations of LLMs for this task. 
In order of priority, the LLM identified 5 methods for relocation (see \circled{2}), including \code{execute}, \code{parse}, \code{optimizedPlan}, and \code{analysePlan}, all of which rightly belong in \code{Esqlsession} and were never moved by developers. Notably, the LLM did successfully identify \code{resolvePolicy} as a candidate for refactoring, showing its ability to detect semantically \emph{misplaced} methods. Despite  success, the LLM recommended other methods before \code{resolvePolicy}. A developer would need to filter out many irrelevant suggestions before arriving at a useful one.

% Given the huge pre-training of LLMs, this impressively shows that LLMs are able to overcome limitations of smaller models that are custom-trained for specific refactorings. %traditional static analysis techniques in identifying refactoring opportunities.
% However, this also exposed significant LLM limitations. 

After identifying that the method \code{resolvePolicy} is misplaced, {a tool must find a suitable target class to move the method into}. While the LLM was able to recommend the correct target class, it also responded with (i) two target classes (i.e., \code{Resolution}, \code{ActionListener}), which are {plausible target classes,}
% \todo{mechanically feasible to move the method into}
but are not the best-fit semantically for the method; (ii) two hallucinations, i.e., classes that do not exist  (i.e., \code{PolicyResolutionService}, \code{PolicyUtils}) as the LLM lacks project-wide context.
% \todo{@Abhiram: please clarify for the first two classes, we initially say they are incompatible but then we say they are feasible; this can be confusing.}
% For this task, the LLM suggested multiple candidate target classes: (1) the appropriate target class (\code{PolicyResolver}).
% This shows that the LLM lacks of awareness of an actual project's structure. 
% Second, the LLM suggested moving the method to \code{Section} (\circled{3} in \cref{fig:motivatingExample}). 
% Although \code{Section} is indeed a valid class in the project, this suggestion reveals a critical flaw in the LLM's reasoning. The \code{Section} class represents a structural element of the API, with responsibilities distinct from utility operations like \code{String} manipulation. 
% This misalignment demonstrates the LLM's inability to determine appropriate class responsibilities. 
{A naive approach to address the LLM's lack of project-wide understanding is to prompt it with the entire codebase. However, this is currently impractical due to the LLM's context size limitations and inability to efficiently handle long contexts \citep{liu2023lost} (even though context limits continuously increase). Even state-of-the-art LLMs can't process large projects like {\em Elasticsearch} in a single prompt without truncating crucial information.}
% \shorten{Moreover, as context capacities expand, processing an entire project at once remains challenging due to increased computational complexity and reduced reasoning effectiveness \citep{liu2023lost}. }
Moreover, the processing cost of large inputs with commercial LLM APIs is prohibitive.

These experiments reveal both the strengths and limitations of LLMs for \mm refactoring. On the positive side, LLMs show proficiency in generating multiple suggestions and demonstrate an ability to identify methods that are semantically misplaced. However, they also exhibit significant limitations, including difficulty in suggesting appropriate target classes, and a high rate of irrelevant or infeasible suggestions. These limitations underscore the need for caution when relying on LLM-generated refactoring recommendations{, and the need for a tedious manual analysis. 
% Developers attempting to leverage LLMs for refactoring would face a laborious and error-prone process.
Developers need to manually collect and re-analyze the suggestions, verify the suitability of each method for relocation, prompt the LLM again for suitable target classes,} and meticulously identify and filter out hallucinations such as non-existent classes and methods that are impossible to move. In the example~(\Cref{fig:motivatingExample}), a developer needs to sift through 5 candidate methods and, for each method, understand if any of the 5 or more proposed target classes are adequate. The developer analyzes 20+ $<$method, target class$>$ pairs before finding one they agree with. 

This example motivates our approach, \tool, which significantly streamlines the refactoring process by (1) utilizing semantic relevance to {find} candidate methods that are the least compatible with the host class, (2) employing static analysis to validate and filter suggestions, and (3) leveraging LLMs to prioritize only valid recommendations. For the example above, \tool expertly recommends as the top candidate moving \code{resolvePolicy} to \code{PolicyResolver}. \tool liberates developers so they can focus on the creative part. 
Rather than sifting through {many} invalid {recommendations}, developers use their expertise to examine a {few} high-quality{~recommendations}.

%% file: Files/approach.tex
\section{Approach}
\label{sec:approach}
% \shorten{In this section we present the workflow that our novel approach and tool, \tool, uses to automatically recommend and perform correctly the \mm refactoring.}  

\begin{figure*}[th]
    \centering    
    \includegraphics[width=0.8\linewidth,trim=0cm 0cm 0cm 0cm]{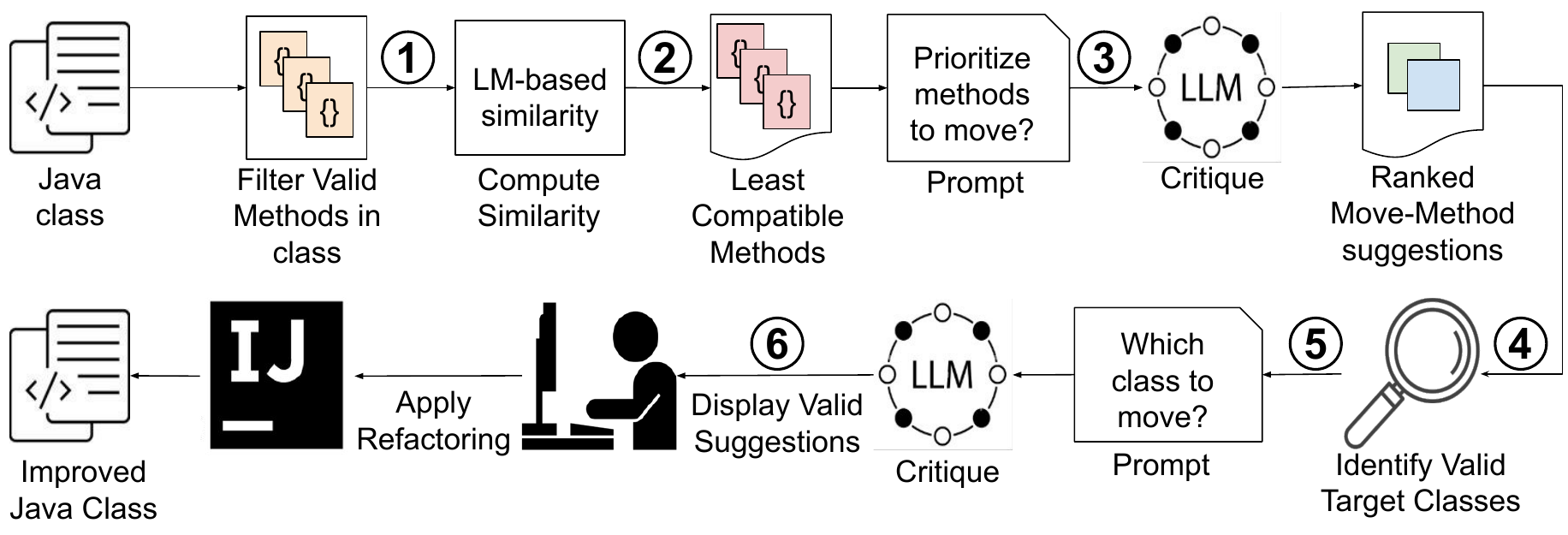}
    \vspace{-10pt}
    \caption{Architecture of \tool. }
    % {\color{red}{Missing (3)}}}
    % \todo{Add numbers to each box, and refer to these specific labeled numbers in the text explaining the workflow.}}
    \label{fig:workflow}
\end{figure*}

{
% method identification
\Cref{fig:workflow} shows the architecture and the steps performed by \tool. First, \tool applies a set of pre-conditions that filter out the methods that cannot be safely moved, such as constructors (\circled{1} in \Cref{fig:workflow}). It then leverages vector embeddings from Language Models to identify methods that are the least cohesive with their host class (\circled{2} in \Cref{fig:workflow}). In~\Cref{fig:motivatingExample}, by comparing the embeddings of \code{resolvePolicy} and \code{EsqlSession} using cosine similarity, \tool detects that this method might be misplaced ($\S$~\ref{subsection:moveMethod}). 
% It identifies other methods in \code{EsqlSession} that might be misplaced based on their low semantic relevance with the host class. 
Next, \tool passes the remaining candidates to the LLM (i.e., the method signature and the class body), which analyzes their 
% implementation, dependencies, and 
relationships with the host class to prioritize the most promising \mm recommendations (\circled{3} in \Cref{fig:workflow}). 

% identify target classes
Once \tool identifies candidate methods, it systematically evaluates potential target classes from the project codebase.
% , considering method accessibility and static/\commentb{instance} constraints ($\S$~\ref{subsection:targetClass}). 
For the \code{resolvePolicy} method, which utilizes the \code{enrichPolicyResolver} field (\circled{1} in \Cref{fig:motivatingExample}), \tool initially identifies several candidate classes, including \code{EnrichPolicyResolver} and \code{PolicyManager}.

% use semantic embedding to narrow down target classes
To narrow down the target classes, \tool calculates relevance scores between the candidate method and each potential target class -- establishing a ranking (\circled{4} in \Cref{fig:workflow}).
% For \code{resolvePolicy}, these scores incorporate multiple factors: semantic coherence with \code{EnrichPolicyResolver}, structural relationships through the \code{enrichPolicyResolver} field reference, parameter compatibility (\code{policyNames}, \code{resolution}), and the overall functional context.
% \tool subsequently ranks candidate classes based on these computed scores to identify the most viable targets (\circled{4} in \Cref{fig:workflow}).
% augment the llm with the retrieved information to select the most suitable class. 
{We label this process as ``refactoring-aware
RAG''.} Then, we feed the LLM with the narrowed-down list of target classes, and ask it to pick the best one (\circled{5} in \Cref{fig:workflow}).
% Using the Retrieval Augmented Generation (RAG) mechanism, we enhance the LLM's input with a prioritized list of target classes and their similarity metrics (\circled{5} in \Cref{fig:workflow}). The LLM processes this enriched context to identify the optimal target class, factoring in code structure, dependencies, and semantic relevance. 
In this case, it correctly selects \code{EnrichPolicyResolver} as the appropriate destination for \code{resolvePolicy}, aligning with the developers' actual refactoring decision (see \circled{3} in \Cref{fig:motivatingExample}). 
% This combination of semantic relevance and LLM-based reasoning allows \tool to produce actionable refactoring suggestions (\circled{6} in \Cref{fig:workflow}). 
Finally, refactoring suggestions (\circled{6} in \Cref{fig:workflow}) are presented to the user, and \tool leverages the IDE's refactoring APIs to safely execute the chosen one automatically. 

Next, we discuss each of these steps and concepts in detail.

\subsection{Important Concepts}

\begin{definition}{(\mm Refactoring\label{def:movemethod})}
{\em A \mm refactoring moves a method \code{m} from a host class \code{H} (where it currently resides) to a target class \code{T}. We define a \mm "$\omega$" as a triplet $(\code{m}, \code{H}, \code{T})$}.

% We denote a single \mm refactoring with the symbol $\omega$, and a set of \mm refactorings with the symbol $\Omega$. Further, for a given \mm refactoring $\omega$, we use the notation $\omega_m$ to denote the method to move ($\code{m}$), $\omega_{\code{H}}$ to denote the host class ($\code{H}$) and $\omega_{\code{T}}$ to denote the target class ($\code{T}$) of the move.}

\end{definition}

% \begin{definition}{(\mm Suggestion\label{def:mmsuggestion})}
% A \mm suggestion is a 
% \end{definition}

\begin{definition}{(\mm Recommendations\label{def:mmsuggestionlist})}
{\em A list of \mm refactoring candidates, ordered by priority (most important at the beginning). We denote this with $\Re$.}

% {\em We define $\Re$ as a tuple of two items ($\Omega$, $\tau$):}
% \begin{enumerate}
%     \item {\em A set of \mm refactorings, $\Omega$,}
%     \item {\em A ranking function, $\tau$, which defines the order of the set $\Omega$. For each item in $\Omega$, $\tau$ emits a number between 1 and $|\Omega|$, which is the position at which it appear in the ranked list,
    
%     $\tau$ must be one-to-one and onto, to ensure that every element in $\Omega$ has a unique ranking. The domain of $\tau$ is $\Omega$ and its range is a natural number $\in [1, |\Omega|]$.

%     Formally, we constrain $\tau: \Omega \rightarrow [1, |\Omega|]$ to be,
    
%     One-to-one: $\forall \omega_{1}, \omega_{2} \in \Omega: \tau(\omega_1) == \tau(\omega_2) \implies \omega_1 = \omega_2$

%     Onto: $\forall i \in [1, |\Omega|]: \exists \omega. \tau(\omega) = i$}
% \end{enumerate}

% We define $\Re$ to be on ordered list of \mm suggestions.{\em This allow us to describe the top-N recommendations made by a system as the first N elements in $\Re$}.

\end{definition}

\begin{definition}{(Valid Refactoring Recommendations\label{def:validrefactoring})}
These are recommendations that do not break the code. 
They are mechanically feasible:
{they successfully pass the preconditions as checked by the IDE, thus resulting in syntactically and semantically equivalent code.} 
We differentiate between moving an instance and a static method:

% 1. An \textbf{Instance Method} can be moved to a type in the host class' fields, or a type among the method's parameters. Several pre-condition checks are necessary to ensure the validity of the \mm suggestion, including:
%     \begin{itemize}    
%         \item Method Movability: Is the method a part of the class hierarchy? \commentb{If not, the method can be safely moved.}
%         \item Access to references: Does the moved method loses access to the references it needs for computation?
%     \end{itemize}

% 2. A \textbf{Static Method} can be moved to almost any class in the project. A valid static-method move is one where the method can still access its references (e.g., fields, methods calls) from the new location.

\begin{enumerate}
    \item An \textbf{Instance Method} can be moved to a type in the host class' fields, or a type among the method's parameters. Several preconditions ensure the validity of the \mm recommendation, including:
    \begin{itemize}    
        \item Method Movability: Is the method a part of the class hierarchy? {If not, the method can be safely moved.}
        \item Access to references: Does the moved method loses access to the references it needs for computation?
    \end{itemize}
    \item A \textbf{Static Method} can be moved to almost any class in the project. A valid static method move is one where the method can still access its references (e.g., fields, methods calls) from the new location.
\end{enumerate}
% Formally, let \( \Omega \) represent the set of all refactoring suggestions generated by a recommendation system. 

% \shorten{We term any \mm suggestion which is not valid to be an invalid suggestion.}

\end{definition}

\begin{definition}{(Invalid \mm Recommendations) \label{def:hallucination}}
% Invalid Move Method Recommendations
% \shorten{LLMs can generate many invalid \mm refactorings that, if executed, would break a software system and produce compile errors. 
We classify the LLM's invalid \mm suggestions as hallucinations and categorize them as follows:
% We identify 
% \todo{three -- but bellow we list only two?}  
% different kinds of hallucinations made by an LLM and define them below:
\begin{enumerate}
    \item {\bf Target class does not exist (H1):} {The LLM comes up with an imaginary class.}
    % Formally, if $\mathbb{C}$ is the set of classes in a project, then for a given \mm suggestion $\omega$, if $\omega_{\code{T}} \notin \mathbb{C}$, we call it a hallucination.
    
    \item {\bf Mechanically infeasible to move (H2):} The target class exists, but a refactoring suggestion is invalid according to definitions in the previous subsection \ref{def:validrefactoring}.

    \item {\bf Invalid Methods (H3)}: The method is a part of the software design, and moving it requires multiple other refactoring changes.
    % to accommodate the move. 
    For example, moving a getter/setter needs to be accompanied by moving the appropriate field.
    
    % \item LLM conflicts with it's previous 
\end{enumerate} 

\end{definition}

% \( C \) the set of syntactically correct suggestions, and \( A \) the set of suggestions that successfully invoke the IDE's refactoring APIs. The set of valid refactoring suggestions \( V \) can then be defined as \( V = C \cap A \), where \( V \subseteq S \), indicating that valid suggestions are both syntactically correct and invoke the APIs successfully, and are a subset of the LLM-generated suggestions.
%\end{definition}

% \begin{definition}{(Executable Refactoring\label{def:executablerefac})}
% We then define Executable Refactoring as a subset of valid refactoring suggestions, denoted by \( E \subseteq V \), where the refactoring can be successfully applied to the method. A valid refactoring suggestion may not be executable due to several reasons, such as conflicts between refactorings that depend on the order of application, or exceptions that the IDE may throw during the refactoring process.
% \end{definition}

\subsection{Identifying Which Method To Move}
\label{subsection:moveMethod}
% We utilize LLMs for \mm refactoring, hypothesizing that their extensive pre-training on vast code repositories and inherent understanding of code naturalness make them well-suited for identifying out-of-place methods. 
{While we believe LLMs have great} potential for suggesting \mm refactorings, directly using LLMs to identify potential methods that may be misplaced within a class is~risky, as it results in invalid \mm recommendations.
% , starting with method-identification, to finding a valid target class.

% First, plausible but invalid refactoring suggestions. Second, the tendency of LLMs to produce hallucinations. Therefore, we decompose the task of identifying move methods into three steps.

{\bf Filter Invalid Candidates via Sanity Checks.}
% \subsubsection{\bf Filtering Invalid Methods via Refactoring Preconditions and Sanity Checks}
\label{subsubsection:methodPrecondition}
Following established refactoring practices~\citep{feTruth,TERRA201819JMove,cui2024three}, we filter 
% In this section, we describe our initial filtering step. {\tool} begins by filtering out methods that fail to meet the basic criteria for relocation through a set of precondition checks. 
% This sanity check eliminates 
methods that are likely already in the correct class. First, {\tool} filters out getter and setter methods, as they cannot be moved without also relocating the associated fields. Next, it excludes methods involved in inheritance chains that can be overridden in subclasses, since moving these would require additional structural changes. It also removes test methods and those with irrelevant content, such as empty bodies or methods containing only comments. 
% \shorten{This filtering ensures that only viable candidates proceed to the next~stages.}

% {\bf Identify Least Compatible Methods via Embedding-Based Analysis.} 

{\bf Identifying Semantically Mismatched Methods.}
% To further refine candidate methods, we use an embedding-based analysis.
{For the remaining candidates, the next step is to find methods that are out of place in their current class. To automate this, we use code embeddings (VoyageAI~\citep{voyageAI}) to capture the semantic meaning of the code. }
An embedding is a vector representation that captures the semantic features of an entity (methods and classes) based on its content and relationships. 
% We leverage VoyageAI embeddings~\citep{voyageAI}, specifically trained on code, as they more effectively capture the semantic relevance of programming constructs. 
We use VoyageAI due to its state-of-the-art performance in code-related tasks. We generate vectors for two inputs: one for the method body and another for the host class, excluding the method body. Excluding the method ensures that the class embedding remains unbiased by the method itself. We then calculate the cosine similarity between these vectors to assess how well each method semantically aligns with its host class.

{\bf Prioritizing Candidates with LLM-driven Reasoning.} 
% \shorten{
% To make the analysis computationally tractable and respect LLM context limitations, we first narrow down candidate methods based on the natural distribution of methods in real-world code. Our analysis of the distribution of the methods across classes reveals a heavy-tailed pattern in which 90\% of the classes contain fewer than 15 methods. This informs our selection of either all methods in the class or the 15 least cohesive methods based on cosine similarity scores, whichever is smaller. We then leverage LLMs to determine if these candidate methods integrate cohesively with the rest of the class. Drawing inspiration from the success of Chain-of-Thought (CoT) reasoning in software engineering tasks~\citep{lecodechain, nextIcml}, we adopt a CoT approach, prompting the LLM to perform structured analyses: evaluate each method's purpose, cohesion, and dependencies, summarize the host class's responsibilities, and assess overall alignment.}
% {To make the analysis computationally tractable and fit the LLM context size, we further narrow  candidate methods based on {LLM's understanding of class level design. 
{Finally, after identifying a pool of semantically misplaced methods, we use an LLM to rank them based on a broader perspective (i.e., class level design).} We use the LLM to rank the existing methods in our suggestion pool. Using Chain-of-Thought (CoT) reasoning, we prompt the LLM to perform a structured analysis: evaluate each method's purpose, cohesion and dependencies, summarize the host class's responsibilities, and assess overall alignment of the method \& class.

% Our prompt is available in our replication package~\citep{ReplicationPackage}.

% This process draws on the LLM's deep understanding of code semantics} and the inherent naturalness of programming constructs, thereby deriving insights that effectively complement the quantitative filtering and embedding-based analysis performed in earlier steps.

\subsection{Recommending Suitable Target Classes}
\label{subsection:targetClass}

After identifying potential methods for relocation, the subsequent task is to determine the most appropriate target classes for these methods. However, this presents a substantial challenge, requiring a comprehensive analysis of the entire codebase. LLMs struggle with such tasks due to their limited context windows. To address this, we employ Retrieval Augmented Generation (RAG)~\cite{lewis2020retrieval}. RAG is a systematic approach designed to retrieve and refine relevant contextual information, thereby augmenting the input to the LLM. In \tool, we efficiently retrieve and augment the model with the most relevant target classes. {Instance methods can be moved to a limited number of feasible classes. Static methods can be moved to almost any class in the project. Thus, we compare the structure and semantics to find the most suitable target classes (described below). In both cases, we rank target classes based on semantic relevance and finally provide a suitable list of target classes to the LLM to allow it to choose the best fit.}
% , informed by a combination of structural filtering and semantic comparison. Structural filtering refers to evaluating characteristics such as package proximity and utility class identification, while semantic comparison involves assessing the meaning and relationships of code elements to determine the most suitable target classes. 
Since we designed the retrieval process to enhance refactoring, we call this ``refactoring-aware RAG'', and we explain it below.

% \subsubsection{\bf Filtering Invalid Target Classes Using \todo{Precondition} Checks}
% {\bf Validating Target Classes via Sanity Checks.} 
% % Once the list of classes with the project is collected, precondition checks are employed to filter out unsuitable target classes (i.e., interfaces and classes with the same method signature) to ensure that it is mechanically feasible to perform the \mm.
% Once target classes are collected, sanity checks filter out unsuitable candidates (i.e., interfaces and duplicate signatures) to ensure mechanical feasibility to perform \mm, following refactoring practices~\citep{feTruth, cui2024three}.
% These precondition checks are derived from both structural and semantic rules, ensuring that a target class can accommodate a method without violating fundamental object-oriented principles such as encapsulation and cohesion. This filtering step guarantees that the resulting set of potential target classes preserves the integrity and coherence of the codebase.

%\commentb{\bf 1. Candidate Filtering by Method Type.} 

\subsubsection{Candidate Filtering by Method Type}
The initial~filtering process differs based on whether the method is an instance or static type, as they possess different relocation constraints.

\paragraph{\textbf{Target Class Retrieval for Instance Methods}}
{As instance methods can be moved to a few suitable classes, we select the method’s parameter types and the host class’ field types as potential destinations. Then, we utilize the IDE's preconditions to retain only valid \mm suggestions.}

\paragraph{\textbf{Target Class Retrieval for Static Methods}}
{We identify potential target classes within the project based on two key aspects: package proximity and utility class identification. Package proximity quantifies structural closeness in the package hierarchy by computing shared package segments (e.g., for org.example.core and org.example.utils, "org" and "example" are shared) normalized by the host package depth, providing an initial structural filter. Utility classes, identified through conventional naming patterns (containing "util" or "utility"), are prioritized as common targets for static methods. These {heuristics} are efficient filters to narrow down the search space of potential target classes.}
{We rank potential target classes based on the above heuristics, with greater importance given to proximity. Our ranking function is:}
% \commentb{We utilize a ranking function to the pottential target classes by considering both their package proximity to the original class and whether they are utility classes, with higher weights given to proximity.}
\begin{equation}
RankingScore(\code{T}) = 2 \cdot \text{proximity}(\code{T}, \code{H}) + \text{isUtility}(\code{T})
\end{equation}
where:
\begin{itemize}
% \item $w_p$ is the weight assigned to package proximity (i.e., $w_p = 2$)
% \item $w_u$ is the weight assigned to utility classes (i.e., $w_u = 1$)
\item $\text{proximity}(\code{T}, \code{H})$ evaluates the package proximity between class $\code{T}$ and the host class $\code{H}$
\item $\text{isUtility}(\code{T})$ is a boolean function that returns 1 if $\code{T}$ is a utility class, and 0 otherwise
\end{itemize}

{Finally, we utilize static analysis from the IDE to validate whether the method can be moved to the potential target class.}

% , with the LLM ultimately making semantically informed decisions about the most suitable targets based on deeper code analysis.
% Let \code{m} be the candidate method for relocation, \code{H} be the host class of \code{m}, $C$ be the set of all classes in the project, and $p_c \subseteq C$ be the set of potential target classes. We formalize the retrieval and ranking as follows:

% \begin{equation}
% p_c = \text{Retrieve}(\code{H}, C)
% \end{equation}
% where $\text{Retrieve}$ is a function that selects potential target classes from the whole project. 

% Then, for each class $t_c \in p_c$, we compute a ranking score $R(c, \code{m})$ that considers both the structural proximity and utility nature of the class:
% \begin{equation}
% R(t_c, \code{m}) = w_p \cdot \text{proximity}(t_c, \code{H}) + w_u \cdot \text{isUtility}(t_c)
% \end{equation}
% where:
% \begin{itemize}
% \item $w_p$ is the weight assigned to package proximity (i.e., $w_p = 2$)
% \item $w_u$ is the weight assigned to utility classes (i.e., $w_u = 1$)
% \item $\text{proximity}(t_c, \code{H})$ evaluates the package proximity between class $t_c$ and the host class of method \code{m}
% \item $\text{isUtility}(t_c)$ is a boolean function that returns 1 if $t_c$ is a utility class, and 0 otherwise
% \end{itemize}

% The resulting $R(t_c, \code{m})$ captures the weighted relevance of each potential target class, considering both structural and functional aspects of the codebase.
% we found that most static methods move to utility and also closer to the original package. 

%\commentb{\bf 2. Semantic Relevance-Based Target Class Ranking.}

\subsubsection{Semantic Relevance-Based Target Class Ranking}
While static analysis offers foundational understanding of valid refactoring opportunities, it often yields a broad set of potential target classes, as it lacks the ability to capture deeper semantic relationships.
% As a result, static analysis alone cannot effectively prioritize or eliminate classes that are only superficially related to the candidate method. 
To augment the results of static analysis, we incorporate a semantic relevance analysis, which {compares} both the content and intent of the candidate method and target classes.
% , aiming to identify deeper semantic connections that static analysis may overlook.
% Our semantic relevance analysis involves two key steps. First, \tool extracts the method body out of the host class. Second, we
To do this, we use {VoyageAI's vector embeddings~\citep{voyageAI}} to compute the cosine similarity between the method body and potential target classes. 
% This helps us to effectively capture semantic relationships between the method and target classes. 
% Formally, the semantic relevance ($rel$) between a method  and a class is computed as follows:
% \shorten{
% \begin{equation} 
% rel(\code{m}, \code{T}) = cosine(embed(\code{m}), embed(\code{T})) 
% \end{equation}
% }
% where embed represents the VoyageAI embedding function. 
% \tool ranks potential target classes for a method based on their relevance scores ($rel$). 
{We sort the target classes by their {cosine similarity scores} in descending order to select the most semantically relevant candidates for the LLM to~analyze. To stay within the LLM's context window, we limit the candidates to those fitting within a 7K token budget --  typically accommodating 10-12 class summaries with their signatures.}

%\commentb{\bf 3. Ranking Target Classes Using LLM.}

\subsubsection{Ranking Target Classes Using LLM}

% \todo{While embeddings capture semantic relevance, they primarily provide vector-based distance metrics. Adding LLM-based analysis enables us to leverage textual features like method names and fields that suggest design intent. }
% After cutting down the number of potential target classes, 
In the final phase, \tool asks the LLM for the best-suited target class, utilizing its vast training knowledge. 
{To avoid context overflow, we create a concise representation of each target class, including its name, field declarations, DocString, and method signatures.} The LLM then takes as input the method to be moved along with these summarized target class representations, returning a prioritized list of target classes.

% By leveraging the language model's extensive training on diverse codebases, this step refines the ranked list, ensuring that the final recommendation is both contextually thorough and semantically accurate.
% \shorten{We formalize this LLM-based ranking process as follows}: 
% \begin{equation} 
% R_{\text{LLM}}(\code{m}) = \text{LLM}\left(\code{m}, p_c^k \right) 
% \end{equation} 
% where: 
% \begin{itemize} 
% \item $R_{\text{LLM}}(\code{m})$ is the final ranked list of target classes for method $\code{m}$ \item $\text{LLM}$ represents the language model's decision-making function 
% \item $\code{m}$ is the candidate method for relocation 
% \item $p_c^k$ is the set of top $k$ potential target classes ranked by $\text{rel}(\code{m}, t_c)$ 
%  \end{itemize}

\subsection{Applying Refactoring Changes}
After compiling a list of \mm recommendations, \tool presents the method-class pairs to developers through an interactive interface, accompanied by a rationale explaining each suggestion. Upon developer selection of a specific recommendation, \tool encapsulates the approved method-target class pair into a refactoring command object. It then executes the command automatically through the IDE's refactoring APIs, ensuring safe code transformation {(moving the method, and changing all call sites and references)}.

%% file: Files/eval.tex
\section{Empirical Evaluation}
\label{sec:eval}

To evaluate {\tool}'s effectiveness and usefulness, we designed a comprehensive, multi-methodology evaluation to corroborate, complement, and expand our findings. This includes~a formative study, comparative study, replication of real-world refactorings, repository mining, user study, and questionnaire surveys. These methods combine qualitative and quantitative data, and together answer four research questions.

\begin{enumerate}[label=\bfseries RQ\arabic*.,wide, labelwidth=!, labelindent=0pt]
\item \textbf{How effective are LLMs at suggesting opportunities for \mm refactoring?} 
% This question assesses vanilla LLMs' ability to identify and suggest \mm refactoring opportunities. We conduct a formative study to understand the strengths and limitations of using LLMs directly for refactoring, \shorten{examining the diversity, feasibility, and} correctness of their suggestions.
{This question assesses Vanilla LLMs' ability to recommend \mm in a formative study examining the quality of LLM suggestions.}

% How effective are LLMs at suggesting refactoring opportunities?
% Formative study of using raw LLM to provide refactoring recommendations: pros and the cons 
% Diversity of refactoring kinds
% How correct are they: (i) check whether suggestion is feasible; (ii) check the correctness of the refactoring performed by the LLM)
% Deeper dive into MoveMethod (LLM hallucinates at the destination class for the moved method)

\item \textbf{How effective is {\tool} at suggesting opportunities for \mm refactoring?} 
We evaluate the performance of \tool against the state-of-the-art tools, \feTruth~\citep{feTruth} and \HMove~\citep{cui2024three} (representatives for DL approaches), and \JMove~\citep{TERRA201819JMove} (representative for static analysis approaches). We used both a synthetic corpus used by other researchers and a new dataset of real refactorings from open-source developers.
% We are comparing on the synthetic corpus against other tools.
% We are replicating the oracle of real refactorings performed by OSS developers
% Sensitivity of the analysis (based on temperature, iterations). How the diversity of recommendations changes with different temperatures and iterations. 
% We compare \tool against other state-of-the-art \mm refactoring tools

% \item \textbf{How do the components of our approach contribute to the overall performance?} 
% This ablation study reveals the impact of different components of \tool. 
% We examine the effectiveness of our method identification technique and compare different code relevance measures (TF-IDF vs. CodeBERT).
% and evaluate the impact of using LLM-based ranking versus similarity-based sorting for target class selection.

% Method Identification: priority based on popularity from reprompts
% (ii) code similarity: cosine vs CodeBert
% Sorting based on ranking from Codebert/cosine, vs LLM choosing the best candidate for Target class 

\item \textbf{What is {\tool}'s runtime performance?} 
This helps us understand \tool's scalability and suitability for integration into developers' workflows. 
% \shorten{We assess its computational efficiency, measuring its runtime performance across various project sizes and complexity levels.} 

\item \textbf{How useful is our approach for developers?} We focus on the utility of \tool from a developer's perspective. We conduct a user study with \numUserParticipants participants with industry experience who used our tool on their own code for a week. 

% \shorten{We analyze their ratings of the quality of recommendations, \tool's usability, and its potential impact on refactoring practices.}

%This question focuses on the practical utility of \tool from a developer's perspective. We conduct a user study with \numUserParticipants graduate students with industrial experience that used \tool on their own authored code for one week. We analyze their feedback regarding the relevance of recommendations, \tool's usability, and its potential impact on developers' refactoring practices.

% User study with students: the rate of acceptance/rejection, the 6-point likert scale on Usefulness, get some quotes from the post-survey (e.g., what they liked the most when using this approach).

% \input{Files/RQs/effectiveness}

\end{enumerate}

\subsection{Subject Systems}
\label{sec:SubjectSystems}
To evaluate LLMs' capability when suggesting \mm refactoring opportunities, we employed two distinct datasets: a synthetic corpus widely used by previous researchers{~\citep{TERRA201819JMove, kurbatova2020recommendation, RMove:2022, cui2024three}} and a new corpus that contains real-world refactorings that open-source developers performed. 
Each corpus comes along with a ``gold set'' $G$ of \mm refactorings that a recommendation tool must attempt to match. We define $G$ as a set of \mm refactorings (see ~\Cref{def:movemethod}) - each containing a triplet of method-to-move, host class, and target class $(\code{m}, \code{H}, \code{T})$.

\subsubsection{Synthetic corpus}
The {\em synthetic corpus} was created by Terra et al.~\citep{TERRA201819JMove} moving different methods \code{m} out of their original/host class \code{H} to a random destination class \code{T}. The researchers then created the gold set as tuples $(\code{m}, \code{H}, \code{T})$, i.e., methods \code{m} that a tool should now move from \code{H} back to its original class \code{T}.
% \shorten{The researchers explained this ensures that method \code{m} is out of place in \code{C'} and can be moved back to original class \code{C}}. 
This dataset moves \emph{only instance methods}; it does not move {\em static methods}. 
%The synthetic nature of this corpus allows for a controlled evaluation environment, enabling us to assess the LLM's performance against a known ground truth. 
This corpus consists of 10 open-source projects.
% see our replication package~\citep{ReplicationPackage}.

% \shorten{(i.e., Ant, Derby, DrJava, JfreeChart, JGroups, JTopen, JUnit, MvnForum, Lucene, Tapestry). On average, one project has 1,574 classes and 13,759 methods, with 207,163 LOC.}

% 
\subsubsection{Real-world corpus}
{As refactorings in the real-world are often complex and messy~\citep{rohit_refactoring_custom}, we decided to} complement the synthetic dataset 
% and provide insights into how closely LLMs resemble the rationale of expert developers in real-world situations, we curated 
with a corpus of \emph{actual} \mm refactorings that open-source developers performed on their projects. {This dataset allows us to determine whether various tools can match the rationale of expert developers in real-world situations.}
We construct this oracle using RefactoringMiner~\citep{RefactoringMiner}, the state-of-the-art  for detecting refactorings~\citep{a_technique_rohit}. 

We took extra precautions to prevent {\em LLM data contamination}, ensuring that the LLMs used by \tool had no prior exposure to the data we tested and could not rely on previously memorized results. With GPT-4’s knowledge cutoff in October 2023, we focused our analysis on repository commits from January 2024 onward.
% \todo{add a sentence summarizing the real-world project sizes, similar with how we did for the synthetic corpus.}

% \shorten{For each commit where RefactoringMiner detected a \mm refactoring, it reports the source and target classes, the original method's signature, and the moved method’s signature.} 
However, many of the \mm reported by RefactoringMiner {are not pure \mm refactorings}, often resulting from residual effects of other refactorings such as \textsc{MoveClass} (where an entire class is relocated to another package) or \textsc{Extract Class} (where a class is split into two, creating a new class along with the original) ~\citep{Novozhilov_Move}. {These impure \mm operations require additional changes to the code's structure, to facilitate them (e.g. creating a new class, moving fields, etc.). To deal with the limited number of data points from \mm refactorings  detected by RefactoringMiner, we leveraged {`Extract and Move Method'} refactoring detected by RefactoringMiner (a composite refactoring in which developers first extract a new method and then move it to a different class). {Empirically, we found that detected `Extract and Move Method's resulted in a greater number of pure refactorings than detected \mm. 
% This can be attributed to two main reasons: first, `the Extract and Move Method' refactoring is frequently performed in practice. Second, it is likely that small sections of a methods are out of place, rather than the entire method, and performing the extract highlights the need for move.} 
This can be attributed to the reason that small sections of a methods are out of place, rather than the entire method, and performing the extract highlights the need for move. Moreover, the `Extract and Move Method' refactoring is frequently performed in practice.}
We constructed our dataset by executing the extract method (within the IDE), and generating an intermediate version of the code for tools to analyze and generate recommendations. Then, we performed further analysis to validate that the target method can in-fact be moved to the target-class. }
{To achieve this,} {we applied a process similar to prior work~\citep{feTruth, cui2024three}}: first, we verified that both the source and target classes existed in both versions of the code (i.e., at the commit head and its previous head). {We removed test methods, getters, setters, and overriden methods (they violate preconditions for a \mm).} For instance methods, we then checked if the method was moved to a field in the source class, to a parameter type, or if the signature of the moved method contained a reference to the source class. Starting from the instances detected by RefactoringMiner, we curated a dataset of \totalRealWorldModern verified \mm, with \totalRealWorldModernStatic static methods and \totalRealWorldModernInstance instance methods -- on 12 popular open-source projects. 
% ~\citep{ReplicationPackage}.
% \shorten{(i.e., {Elasticsearch, Spring-framework, Selenium, Ghidra, Vue-pro, Halo, Kafka, Graal, Redisson, selenium, Apache Flink, Spring-boot})}. 
On average, each project contains 8743 classes and 66306 methods spanning {1032344 LOCs}. This oracle enables an evaluation of \tool on authentic refactorings made by experienced developers. 

% \commentb{After excluding repositories without meaningful business logic implementations (e.g., competitive programming), refactorings that are not inherently supported by the IDE because of xyz, and false positive refactorings.} We ended up with a set of {12} projects (i.e., {Elasticsearch, Spring-framework, Selenium, Ghidra, Vue-pro, Halo, Kafka, Graal, Redisson, selenium, Apache Flink, Spring-boot}) and \totalRealWorldModern refactoring data points. On average, a project contains 8,743 classes and 66,306 methods spanning \todo{1,032,344 LOCs}. This real-world oracle enables an evaluation of the LLM’s performance on authentic refactorings made by experienced developers.

% \input{Tables/synthetic_benchmark}

%% file: Files/results.tex
\input{Files/RQs/rq1_result}
\input{Files/RQs/rq2_result}

\input{Files/RQs/rq3_result}
\input{Files/RQs/rq4_result}

%% file: Files/RQs/rq1_result.tex
% RQ1
\subsection{Effectiveness of LLMs (RQ1)}

% Both corpora present an oracle of \mm refactorings: moving a certain method to the class where it actually belongs inside. We call this set of \mm refactorings (see \ref{def:movemethod}) as a "gold-set" ($G$), similar to ~\citep{TERRA201819JMove}. 

%{\bf Evaluation Metrics.} 

\subsubsection{Evaluation Metrics}

Using these datasets, we evaluated the recommendations made by the vanilla LLM and identified the hallucinations, as defined in \Cref{def:hallucination}: {H1 -- target class does not exist in the project; H2 -- moving the method to the target class is mechanically infeasible; and H3 -- violating preconditions in \Cref{subsubsection:methodPrecondition}}.

% the LLM's self-consistency test, i.e., the \emph{critique LLM} rejects recommendations made by the \emph{suggester LLM}. 
%We report these numbers as well, as they are methods that are falsely identified as candidates.
% Finally, we also computed the recall of using the vanilla LLM. 
% \todo{Abhiram: improve this text}

% \todo{Abhiram: : Can you write a few words about the eval metric here?}

\subsubsection{Experimental Setup}

We use the vanilla implementation of GPT-4o, a state-of-the-art LLM from OpenAI \citep{ChatGPT}. {We used a version that was released on May 13, 2024.} {While {\tool} is model agnostic (i.e., we can simply swap different models)}, we chose GPT-4o because researchers~\citep{PomianICSME2024,pycraft}
 show that it outperforms other LLMs when used for refactoring tasks. GPT-4o is also widely adopted in many {software-engineering tools}~\citep{chiang2024chatbot,arenaLeaderBoard,Cursor,copilot}. We designed our experimental setup to assess the model's inherent capabilities in understanding and recommending \mm without additional context or task-specific tuning. 
 {We formulated a prompt where we provided the source code in a given host class and asked the LLM which methods to move and where}. 
 % \shorten{We formulated a prompt where we provided the source code in a given Host class and asked the LLM which methods are more appropriately placed in other classes and in which classes to move such methods}. 
 % The exact prompt is in our companion webpage~\citep{ReplicationPackage.
 We set the LLM temperature parameter {to 0 to obtain deterministic results}.

 %structure is shown in Listing~\ref{lst:vanilla_mm_prompt}.
% \begin{figure}[h!]
% \centering
% \promptstyle
% \begin{lstlisting}[escapechar=|,caption=Vanilla Prompt to suggest Move Method Refactoring, label=lst:vanilla_mm_prompt]
% Given the following Java class:

% [SOURCE_CLASS_CODE]

% Analyze this class and suggest any methods that would be more appropriately placed in another class. For each suggestion, provide the target class. 
% \end{lstlisting}
% \end{figure}
% striking a balance between creativity and consistency in the generated suggestions. This setting allows for some variability in the model's outputs while maintaining a reasonable level of determinism.
For each host class in our Gold sets (synthetic and real-world), we submitted prompts to the LLM and collected its recommendations. 
% \shorten{We then evaluated them against the ground truth. Moreover, we conducted a qualitative analysis of the LLM's explanations and target class suggestions to gauge the depth of its understanding and possible hallucinations.} 

\input{Tables/rq1_hallucinations}

%{\bf Results.} 

\subsubsection{Results}
\Cref{table:rq1_hallucinations} illustrates the distribution of valid suggestions and different types of hallucinations produced by the vanilla LLM for both synthetic and real-world datasets. 
{We observed a prevalence in all three types of hallucinations: H1, H2, and H3 (as defined in \Cref{def:hallucination})}
% \shorten{We observed three main types of hallucinations: Non-existent target classes (H1), where the LLM suggested moving methods to classes that don't exist; Unfeasible target classes (H2), where the proposed refactorings would break compilation due to inaccessible target classes; and Incorrect method identification (H3), where the LLM mistakenly flagged well-placed methods for relocation.}
{\em Crucially, actuating any of these hallucinations would lead to broken code, compilation errors, or degraded software design.}

{In both the synthetic and real-world dataset, a mere \llmFinalUsefullPercentageExtendedCorpus (142/723 and 267/1293 respectively) were valid. The overwhelming 80\% were hallucinations.}
% \shorten{In the synthetic dataset, comprising 723 total suggestions, a mere \llmFinalUsefullPercentageSyntheticCorpus (142) were valid. The overwhelming 80\% (581) were hallucinations, with H1 accounting for 50.1\% (362), H2 for 23.2\% (168), and H3 for 7.1\% (51) of all suggestions. 
% The real-world dataset
% % % , while slightly better, still
% also presented significant challenges. Once again, out of \totalmmAssistSuggestions total suggestions, only \llmFinalUsefullPercentageExtendedCorpus were valid. The \totalInvalidSuggestionsRate hallucinations were distributed as follows: H1 comprised \typeOneHallucinationRate (\typeOneHallucination), H2 \typeTwoHallucinationRate (\typeTwoHallucination), and H3 \typeThreeHallucinationRate (\typeThreeHallucination) of all suggestions.}
These findings underscore the impracticality of using vanilla LLM recommendations for \mm without extensive filtering and validation. For every valid recommendation, a developer would need to sift through and discard 3-4 invalid ones -- which may introduce critical errors if implemented. This undermines the potential time-saving benefits of automated refactoring and introduces significant risks of introducing bugs or degrading code quality.

% \input{Tables/rq1_table}
% \todo{add text: I love this \Cref{fig:rq1_hallucinations} as it shows very visually that the majority of recommendations are hallucinations and they further get subdivided in groups. How can we also add either absolute numbers on these buckets, or some percentages so that people can also see the actual numbers?}

% \resultbox{
% \shorten{LLMs excel at generating \mm recommendations, yet only \llmFinalUsefullPercentageExtendedCorpus of these suggestions are useful.
% }}

%% file: Tables/rq1_hallucinations.tex
% \#suggestions &
% \#$Hall_C$ &
% \#$Hall_{mech}$ &
% \#critique reject
\begin{table}[ht]
\centering
\scriptsize
\caption{Different kinds of hallucinations from Vanilla LLM}
\resizebox{\columnwidth}{!}{%
\begin{tabular}{|l|l|l|l|l|}
\hline
\textbf{Corpus} & \textbf{\# R} & \textbf{\# H1} & \textbf{\# H2} & \textbf{\# H3} \\  
\hline
Synthetic (235) & 723 & 362 & 168 & 51 \\
\hline
Real-world (\totalRealWorldModern) & \totalmmAssistSuggestions & \typeOneHallucination & \typeTwoHallucination & \typeThreeHallucination \\
\hline
\end{tabular}%
}
\vspace{3pt}
\scriptsize{R: Recommendations, H1: Hall-class, H2: Hall-Mech, H3: Invalid Method.}
\label{table:rq1_hallucinations}
\end{table}

% \begin{figure}[h]
%     \centering    
%     \includegraphics[width=0.6\linewidth,trim=0cm 0cm 0cm 0cm]{Figures/RQ1_Hallucination.pdf}
%     \vspace{-15pt}
%     \caption{Different kinds of hallucinations made by the Vanilla LLM} 
%     \vspace{9pt}
%     \label{fig:rq1_hallucinations}
% \end{figure}

%% file: Files/RQs/rq2_result.tex
\subsection{Effectiveness of \tool (RQ2)}

To evaluate \tool's effectiveness, we compared it against state-of-the-art \mm recommendation tools.

%{\bf Baseline Tools.} 

\subsubsection{Baseline Tools}

We directly compare with the best-in-class tools:
{
\JMove~\citep{TERRA201819JMove} is a state-of-the-art static analysis tool, \feTruth~\citep{feTruth} is the former best in class tool that uses ML/DL, and
\HMove~\citep{cui2024three} is a recently introduced state-of-the-art tool that uses graph neural networks to {generate suggestions} and LLM to check refactoring preconditions. \HMove has been shown to outperform all previous tools.} We also compare with the Vanilla-LLM (GPT 4o), which represents the standard LLM solution (without using \tool's enhancements). We went the extra mile to ensure the most fair comparison: we consulted with \HMove, \feTruth, and \JMove authors to ensure the optimal tool's settings and clarified with the authors when their tools did not produce the expected results. We are grateful for their assistance. {Further, we also replicated the recall numbers presented in each tool's paper, validating our experimental setup.}

\subsubsection{Evaluation Metrics}

For evaluation, we employ recall-based metrics, following an approach similar to that used in the evaluation of JMove~\citep{TERRA201819JMove}, a well-established tool in this domain. 
{
In our setting, recall is a more suitable metric (against precision) because it measures how many relevant recommendations are retrieved, avoiding the need for subjective judgments about whether a recommendations is a false positive.
% Recall is a more suitable metric in our setting because it captures the extent to which all relevant suggestions are retrieved rather than relying on potentially subjective judgments of whether a suggestion is truly a false positive.
% Recall is a more suitable metric for evaluation because it provides a measure of performance that is not subject to individual interpretation, unlike precision. We believe, that only the authors of the code can determine whether a suggestion is a false-positive.
Furthermore, to provide actionable recommendations and avoid overwhelming the developer, we use recall@k, which returns a small number of recommendations (k).
% that are likely to be useful to the practitioner. 
Furthermore, recall@k is similar to precision by evaluating the quality of a limited set of recommendations. 
% rather than all of them.
}
% For refactoring recommendation solutions that aim to be used by industry practitioners, recall@k (where k is a small number so that practitioners do not sift through several candidates) is a more appropriate metric. Moreover, it is more fair than precision as it is \emph{not subjective}. Given that a tool can produce reasonable recommendations, but only the authors of those software systems can accurately determine whether those recommendations are (un)reasonable, we cannot calculate a true precision. But we can calculate a true recall as we have a reliable Gold set (G).

We present recall for each phase of suggesting the move-method refactoring: first, identifying that a method is misplaced, no matter the recommened target class ($Recall_M$); second, identifying a target class for the misplaced method ($Recall_C$); third, identifying the entire chain of refactoring: selecting the right method and the right target class ($Recall_{MC}$).
% \commentb{If a tool correctly identifies the oracle method as a candidate for a move (no matter the recommended target class), it counts towards $Recall_M$. Starting from correctly identified target methods, the cases where a tool found the correct target class, counts towards $Recall_C$. Only the cases where the \mm refactoring exactly matches the oracle counts towards $Recall_{MC}$.}
% \begin{enumerate}
%     \item Identifying that a method is misplaced. ($Recall_M$)
%     \item Identifying a target class for the misplaced method. ($Recall_C$)
%     \item Identifying the entire chain of refactoring: selecting the right method and the right target class. ($Recall_{MC}$)
% \end{enumerate}

For a recommendation list $\Re$ (see \ref{def:mmsuggestionlist}), we define  $Recall_{M}$, $Recall_{C}$ and $Recall_{MC}$ as follows:
% $Recall_{M} = \dfrac{|\Re_M|} {|G|}$, $Recall_{C} = \dfrac{|\Re_M \cap G|}{|\Re_M|}$, and$\hspace{2cm}Recall_{MC}=\dfrac{|(\Re \cap G)|} {|G|}$
\[
\scalebox{1.0}{$
    \mathbf{Recall}_\mathbf{M} = \frac{|\Re_{M}|}{|G|}, 
  % \quad
  \hspace{0.5em}
  \mathbf{Recall}_\mathbf{C} = \frac{|\Re \cap G|}{|\Re_{M}|},
  % \quad
  \hspace{0.5em}
  \mathbf{Recall}_\mathbf{MC} = \frac{|\Re \cap G|}{|G|}
$}
\]

% We define the set of \mm refactorings in $\Re$ where the method was correctly identified as follows:
Where $\Re_M$ is the subset of $\Re$ containing refactorings whose method components match those in the ground truth set $G$. Formally, we define $\Re_M$ as follows: 

\hspace{1cm} $\Re_M =\{g | g \in G \wedge \exists (g_m, g_c, *) \in \Re\} $

% Consequently, we define $Recall_{M}$, $Recall_{C}$ as follows:

For each recall metric, we calculate Recall@k for the top $k$ recommendations, where k $\in$ \{1, 2, 3\}. 
% Recall@k considers only the top-ranked suggestions. 

% \commentb{This is a straightforward process in the synthetic corpus, as there is only one method that needed to be moved from a single source class. However, in the real-world corpus, we mined cases where multiple \mm refactorings were applied to a single source class.
% \\
% Accordingly, we adjust the definitions of $Recall_M$, $Recall_C$ and $Recall_{MC}$. If there are two methods moved from a source class, then the previous definition of recall@1 value would be capped at 50\%, as there are two \mm in the gold set, but there can only be one \mm recalled at most. This is harsh on the tools, and so we change the definition of recall@N as follows:
% }

% We compute recall@N as follows: recall for the top-N ranked suggestions.

% For all of the above, we compute recall@1, recall@2 and recall@inf.

%Now we describe how we conducted the comparative study between \tool and three other tools. We use the state-of-the-art \mm recommendation tools as our baseline: \JMove \citep{TERRA201819JMove} -- which utilizes structural and semantic information and is the best-in-class for static analysis approaches, and \feTruth~\citep{feTruth} which is the best in class for DL/ML approaches. We also compare with the Vanilla-LLM (GPT 4.0), which represents the standard LLM solution (without using \tool's enhancements). 

%{\bf Experimental Setup.} 

\subsubsection{Experimental Setup}

We trigger all tools on each host class from the gold set (both synthetic and the real-world dataset). {To account for the inherent non-determinism in LLMs, we ran the vanilla LLM with multiple temperature values (0, 0.5, 1) for five runs each. Additionally, we ran \tool three times. We report the average and standard deviation for both vanilla LLM and \tool.}
{Considering the number of entries in the datasets, given that \JMove can take a long time to run (12+ hours on a large project), we cutoff its execution after 1 hour}. {\HMove takes both the method and candidate target class as input, and returns a probability score indicating whether to move the method. As a result, it becomes impractical to recommend moving static methods, because there can be thousands of (method, target-class) pairs -- each taking a minute on average to classify.
% 56 seconds on average.
% The resulting complexity is \(O(m \times C)\). 
For example, to recommend moving a single static-method in the Elasticsearch project with 21615 target classes would require 
% $21615 \times 56 
$\approx 14$ days to process. Thus, we limit \HMove to instance move method recommendations.
{We do not penalize any tool for running out of time – we do not compute recall rates on those data points. For example in Table 3, \JMove only finished computing for 24 out of 70 Large Classes, so we only report \JMove’s Recall for those 24 Classes (i.e., we do not compute nor report a zero recall for the 46 Classes it ran out of time). 
}
} 

% All the tools generate a ranked list of possible target classes (\code{T}) for each candidate method (\code{m}). 
{All the tools generate a ranked list of \mm suggestions.}
We compared these {suggestions} against the ground truth to calculate recall for each tool using the evaluation metrics presented earlier { -- $Recall_M$, $Recall_C$, and $Recall_{MC}$.}
\input{Tables/rq2_synthetic_comparison_table}

%{\bf Results on the Synthetic Corpus.} 

\subsubsection{Results on the Synthetic Corpus}

Table~\ref{table: rq3table} shows the effectiveness of \tool and baseline tools on the synthetic dataset. {\tool demonstrates superior performance} across many of the recall metrics compared to other tools, especially $Recall_{MC}@1$ ({1.7x to 30}x improvement) -- the most comprehensive measure assessing both correct method and target class identification. 
% \shorten{Most importantly, it shows an increase of 17-71\% compared to all baselines in $Recall_{MC}@1$, which is the most comprehensive measure as it captures both the correct identification of misplaced methods and the accurate suggestion of target classes.} 
While \JMove exhibits high accuracy in target class identification ($Recall_C@1$ = 97\%), it shows limitations in method identification. {\HMove demonstrates comparable performance as \JMove in identifying misplaced methods (i.e., \(Recall_M\)) but does not match \JMove's ability to recommend the correct target class (i.e., \(Recall_C\)). Consequently, its combined \(Recall_{MC}\) is lower than \JMove.}
% \shorten{indicating that while \HMove occasionally identifies methods more effectively, it still does not achieve the end-to-end accuracy attained by the top-performing approaches.}
Interestingly, \feTruth achieves perfect $Recall_C$ but extremely low $Recall_M$ (2\% -- 3\%) despite being very prolific in recommending as many as 67 methods to be moved from a class.
% (see Replication Package~\citep{ReplicationPackage}).  
When it does correctly identify a method, it accurately suggests the target class. 
{However, because it rarely identifies the misplaced methods themselves, the overall $Recall_{MC}$ remains low.} We confirmed this paradox with \feTruth authors. 
% However, this high $Recall_C$ is less meaningful, thus resulting in a low overall $Recall_{MC}$ due to the low number of correctly identified misplaced methods.
{Interestingly, the Vanilla-LLM shows comparable performance to \tool in method identification ($Recall_M$). This could be because the LLMs have been pre-trained on the synthetic dataset, and might have memorized their responses -- thus becoming necessary to use an LLM uncontaminated dataset (see next). 
% Additionally, many of the oracle methods in the synthetic dataset can easily be identified as misplaced -- there are many getters, setters, test-cases in the wrong class.
}
% but fails to suggest the target class accurately. 
% \shorten{In contrast to all previous tools, \tool's consistent high performance across all metrics, especially in ($Recall_{MC}@3$ = 80.4\%), highlights the effectiveness of our approach in combining LLM capabilities with static analysis and semantic relevance. }

%{\bf Results on the Real-world Corpus.} 

\subsubsection{Results on the Real-world Corpus}

With the real-world dataset performed by open-source developers, we found a wider difference in performance.
 % based on changing two variables
First, we distinguish between cases when the \mm target was an instance or static method { -- shedding light on the effectiveness of \tool in different usage scenarios.} 
 Second, we distinguish between small and large classes based on their method count. Our analysis of the real-world oracle reveals a heavy-tail distribution -- we label classes with fewer than 15 methods (90th percentile across all projects) {as Small Classes (avg. 6 methods/class) and the rest as Large Classes (avg. 48 methods/class)}.
% Second, we distinguish the cases when the refactoring was performed in a small or large class \todo{FIxMe: say how we define a large class} -- thus we shed light on \todo{X?}. 

\input{Tables/rq2_real_instance}

\input{Tables/rq2_real_comparison_table}

Tables ~\ref{table: rq2_real_instance} and ~\ref{table: rq2_real} summarize our results for  instance and static method moves, respectively. {Since \JMove could not finish running sometimes
% on the entire dataset 
and \HMove sometimes failed with a syntax error due to unsupported Java language features, we note the number of completed entries in parenthesis}. For instance methods in Small Classes, \tool achieved {2.4x to 4x ($Recall_{MC}@3$)} higher recall compared to baseline tools.
% While recalling instance \mm, 
For Small Classes, we noticed that our performance
% on small classes 
was comparable to the synthetic dataset, while for other tools dropped significantly. Notably, our $Recall_C$@3 was \rwisRecallCThree, which can be attributed to the performance of the LLM in picking the suitable target classes. 
However, we observed a performance degradation in all tools when identifying \mm opportunities in large classes. This happens because  large classes are more prone to significant technical debt, and there are many candidate methods that can be moved -- thus it is harder to pick the proverbial ``needle from the haystack''. 
% \shorten{Since we are not the developers of these classes, we are not the best to judge the merit of each recommendation, thus we rely on whether the tool matched exactly what the open-source developers refactored.}

% \todo{how many recomendations from jmove, fetruth per class, on avg. How many total recomendations. to validate that we run the tools correctly.}

However, the differences are more nuanced when we evaluate \tool on static methods: we find that our $Recall_C$ drops significantly. This is because the scope of moving static methods is massive - they can be moved to (almost) any class in the project. For large projects like Elasticsearch, this means picking the right target class among 21615 candidates. The real-world oracle contains on average 8743 classes per project.
% \todo{Also say how many classes are in the other projects used in our corpus of real-world -- this further highlights that the problem is more challenging.}
This shows that recommending which static methods to move is a much harder problem than recommending instance methods, as a tool should analyze thousands of classes to find the right one. This could explain why prior \mm tools do not give recommendations for moving static methods. 
As we are the first ones to make strides in this harder problem, we hope that by contributing this dataset of static \mm to the research community, we stimulate growth in this area.

% \Cref{table: rq1_hallucinations} shows the different kinds of hallucinations that LLM makes. \todo{We notice that...}

% \input{Tables/rq2_table}

% \resultbox{
% \shorten{\tool outperforms all baselines by 2.5x
% }}

%% file: Tables/rq2_synthetic_comparison_table.tex
\begin{table*}[t]
\centering
\caption{Recall rates of \tool on the synthetic corpus of 235 refactorings~\citep{TERRA201819JMove} that moved instance methods. $Recall_M$ = identify the method, $Recall_C$ = identify the target class for a method, $Recall_{MC}$= identify the method\&target class pair }
\begin{tabular}{|l|lll|lll||lll|}
\hline
\multicolumn{1}{|c|}{\multirow{2}{*}{Approach}} & \multicolumn{3}{c|}{$Recall_M$} & \multicolumn{3}{c|}{$Recall_C$} &  \multicolumn{3}{c|}{$Recall_{MC}$}                                                                \\ \cline{2-10} 
\multicolumn{1}{|c|}{}                          & \multicolumn{1}{c|}{@1} & \multicolumn{1}{c|}{@2} & \multicolumn{1}{c|}{@3} & \multicolumn{1}{c|}{@1} & \multicolumn{1}{c|}{@2} & \multicolumn{1}{c|}{@3} & \multicolumn{1}{c|}{@1} & \multicolumn{1}{c|}{@2} & \multicolumn{1}{c|}{@3} \\ \hline

\JMove                                     & 
\multicolumn{1}{l|}{41\%} & 
\multicolumn{1}{l|}{43\%} & 
\multicolumn{1}{l|}{43\%} & 
\multicolumn{1}{l|}{97\%} & 
\multicolumn{1}{l|}{97\%}     & 
\multicolumn{1}{l|}{97\%} &
\multicolumn{1}{l|}{40\%}     & 
\multicolumn{1}{l|}{42\%} & 
\multicolumn{1}{l|}{42\%} 
\\ \hline

\feTruth                                     & 
\multicolumn{1}{l|}{2\%} & 
\multicolumn{1}{l|}{3\%} & 
\multicolumn{1}{l|}{3\%} & 
\multicolumn{1}{l|}{\bf 100\%} & 
\multicolumn{1}{l|}{\bf 100\%}     & 
\multicolumn{1}{l|}{\bf 100\%} &
\multicolumn{1}{l|}{2\%}     & 
\multicolumn{1}{l|}{3\%} & 
3\%
\\ \hline

\HMove & 
\multicolumn{1}{l|}{ {31\%}} & 
\multicolumn{1}{l|}{ {37\%}} & 
\multicolumn{1}{l|}{ {40\%}} & 
\multicolumn{1}{l|}{{32\%}} & 
\multicolumn{1}{l|}{{37\%}}     & 
\multicolumn{1}{l|}{{39\%}}  &
\multicolumn{1}{l|}{ {21\%}}     & 
\multicolumn{1}{l|}{ {24\%}} & 
{ {26\%}} 

\\ \hline

Vanilla-LLM & 
\multicolumn{1}{l|}{71$\pm$2\%} & 
\multicolumn{1}{l|}{75$\pm$1\%} & 
\multicolumn{1}{l|}{79$\pm$2\%} & 
\multicolumn{1}{l|}{70$\pm$1\%} & 
\multicolumn{1}{l|}{70$\pm$1\%}     & 
\multicolumn{1}{l|}{70$\pm$1\%} &
\multicolumn{1}{l|}{53$\pm$2\%}     & 
\multicolumn{1}{l|}{55$\pm$2\%} & 
57$\pm$2\% \\ \hline

\tool & 
\multicolumn{1}{l|}{\bf 72$\pm$1\%} & 
\multicolumn{1}{l|}{\bf 79$\pm$0\%} & 
\multicolumn{1}{l|}{\bf 80$\pm$0\%} & 
\multicolumn{1}{l|}{91$\pm$1\%} & 
\multicolumn{1}{l|}{97$\pm$0\%}     & 
\multicolumn{1}{l|}{98$\pm$1\%}  &
\multicolumn{1}{l|}{\bf 67$\pm$0\%}     & 
\multicolumn{1}{l|}{\bf 73$\pm$0\%} & 
{\bf 75$\pm$0\%} 

\\ \hline

\end{tabular}
\label{table: rq3table}
\end{table*}

%% file: Tables/rq2_real_instance.tex
\begin{table*}[]
\centering
\caption{Recall rates on \totalRealWorldModernInstance instance methods moved by {OSS developers in 2024}. First column shows the number of small or large classes in the oracle. $Recall_M$ = identify the method, $Recall_C$ = identify the target class for a previously identified method to be moved, $Recall_{MC}$= identify the method\&target class pair.  }
% {For \feTruth we report recall rates only for a subset of \feTruthProcessedRealWorld refactorings, for the remaining refactorings \feTruth did not recommend any refactorings within the 1-hour time-limit.}}
\begin{tabular}{|l|l|lll|lll||lll|}
\hline
\multicolumn{1}{|c|}{\multirow{2}{*}{Oracle Size}} & 
\multicolumn{1}{c|}{\multirow{2}{*}{Approach}} & 
\multicolumn{3}{c|}{$Recall_M$} & \multicolumn{3}{c|}{$Recall_{C}$} &  \multicolumn{3}{c|}{$Recall_{MC}$}                                                                \\ \cline{3-11} 
\multicolumn{1}{|c|}{} &\multicolumn{1}{c|}{}                          & \multicolumn{1}{c|}{@1} & \multicolumn{1}{c|}{@2} & \multicolumn{1}{c|}{@3} & \multicolumn{1}{c|}{@1} & \multicolumn{1}{c|}{@2} & \multicolumn{1}{c|}{@3} & \multicolumn{1}{c|}{@1} & \multicolumn{1}{c|}{@2} & \multicolumn{1}{c|}{@3} \\ \hline

\multirow{3}{*}{SmallClasses (38)} &
\JMove (19)  & 
\multicolumn{1}{l|}{{5\%}}     & 
\multicolumn{1}{l|}{{5\%}}  & 
\multicolumn{1}{l|}{{5\%}}  & 
\multicolumn{1}{l|}{{0\%}}  & 
\multicolumn{1}{l|}{{0\%}}  & 
\multicolumn{1}{l|}{{0\%}}  & 
\multicolumn{1}{l|}{{0\%}}  & 
\multicolumn{1}{l|}{{0\%}}      & 
{0\%}                       \\ \cline{2-11} 

& \feTruth   & 
\multicolumn{1}{l|}{20\%}     & 
\multicolumn{1}{l|}{20\%}  & 
\multicolumn{1}{l|}{20\%}  & 
\multicolumn{1}{l|}{\bf 100\%}  & 
\multicolumn{1}{l|}{\bf 100\%}  & 
\multicolumn{1}{l|}{\bf 100\%}  & 
\multicolumn{1}{l|}{20\%}  & 
\multicolumn{1}{l|}{20\%}      & 
20\%                  \\ \cline{2-11} 

& \HMove (30)   & 
\multicolumn{1}{l|}{\HMOVErwisRecallMOne}     & 
\multicolumn{1}{l|}{\HMOVErwisRecallMTwo}  & 
\multicolumn{1}{l|}{\HMOVErwisRecallMThree}  & 
\multicolumn{1}{l|}{\HMOVErwisRecallCOne} & 
\multicolumn{1}{l|}{\HMOVErwisRecallCTwo}  & 
\multicolumn{1}{l|}{\HMOVErwisRecallCThree}  & 
\multicolumn{1}{l|}{\HMOVErwisRecallMCOne}  & 
\multicolumn{1}{l|}{\HMOVErwisRecallMCTwo}      & 
\HMOVErwisRecallMCThree                  \\ \cline{2-11}

&Vanilla-LLM   & 
\multicolumn{1}{l|}{\rwisRecallMOneVanilla$\pm$3}     & 
\multicolumn{1}{l|}{\rwisRecallMTwoVanilla$\pm$3}  & 
\multicolumn{1}{l|}{\rwisRecallMThreeVanilla$\pm$3}  & 
\multicolumn{1}{l|}{\rwisRecallCOneVanilla$\pm$4}  & 
\multicolumn{1}{l|}{\rwisRecallCTwoVanilla$\pm$4}  & 
\multicolumn{1}{l|}{\rwisRecallCThreeVanilla$\pm$4}  & 
\multicolumn{1}{l|}{\rwisRecallMCThreeVanilla$\pm$3}  & 
\multicolumn{1}{l|}{\rwisRecallMCTwoVanilla$\pm$2}      & 
\rwisRecallMCThreeVanilla$\pm$3                       \\ \cline{2-11} 

&\tool  & 
\multicolumn{1}{l|}{\bf \rwisRecallMOne$\pm$1}     & 
\multicolumn{1}{l|}{\bf \rwisRecallMTwo$\pm$0}  & 
\multicolumn{1}{l|}{\bf \rwisRecallMThree$\pm$0}  & 
\multicolumn{1}{l|}{ \rwisRecallCOne$\pm$1}  & 
\multicolumn{1}{l|}{ \rwisRecallCTwo$\pm$0}  & 
\multicolumn{1}{l|}{ \rwisRecallCThree$\pm$0}  & 
\multicolumn{1}{l|}{\bf \rwisRecallMCOne$\pm$2}  & 
\multicolumn{1}{l|}{\bf \rwisRecallMCTwo$\pm$1}      & 
{\bf \rwisRecallMCThree$\pm$1}                        \\ \hline \hline

\multirow{3}{*}{LargeClasses (70)} &
\JMove ({24})    & 
\multicolumn{1}{l|}{{8\%}}     & 
\multicolumn{1}{l|}{{8\%}}  & 
\multicolumn{1}{l|}{{8\%}}  & 
\multicolumn{1}{l|}{\bf {100\%}}  & 
\multicolumn{1}{l|}{\bf {100\%}}  & 
\multicolumn{1}{l|}{\bf {100\%}}  & 
\multicolumn{1}{l|}{{8\%}}  & 
\multicolumn{1}{l|}{{8\%}}      & 
{8\%}                  \\ \cline{2-11} 

&\feTruth    & 
\multicolumn{1}{l|}{2\%}     & 
\multicolumn{1}{l|}{8\%}  & 
\multicolumn{1}{l|}{12\%}  & 
\multicolumn{1}{l|}{76\%}  & 
\multicolumn{1}{l|}{76\%}  & 
\multicolumn{1}{l|}{76\%}  & 
\multicolumn{1}{l|}{2\%}  & 
\multicolumn{1}{l|}{6\%}      & 
9\%                  \\ \cline{2-11} 

&\HMove (55)    & 
\multicolumn{1}{l|}{\HMOVErwilRecallMOne}     & 
\multicolumn{1}{l|}{\HMOVErwilRecallMTwo}  & 
\multicolumn{1}{l|}{\HMOVErwilRecallMThree}  & 
\multicolumn{1}{l|}{\HMOVErwilRecallCOne} & 
\multicolumn{1}{l|}{\HMOVErwilRecallCTwo}  & 
\multicolumn{1}{l|}{\HMOVErwilRecallCThree}  & 
\multicolumn{1}{l|}{\HMOVErwilRecallMCOne}  & 
\multicolumn{1}{l|}{\HMOVErwilRecallMCTwo}      & 
\HMOVErwilRecallMCThree                  \\ \cline{2-11} 

&Vanilla-LLM   & 
\multicolumn{1}{l|}{\rwilRecallMOneVanilla$\pm$2}     & 
\multicolumn{1}{l|}{\rwilRecallMTwoVanilla$\pm$2}  & 
\multicolumn{1}{l|}{\rwilRecallMThreeVanilla$\pm$2}  & 
\multicolumn{1}{l|}{\rwilRecallCOneVanilla$\pm$8}  & 
\multicolumn{1}{l|}{\rwilRecallCTwoVanilla$\pm$8}  & 
\multicolumn{1}{l|}{\rwilRecallCThreeVanilla$\pm$8}  & 
\multicolumn{1}{l|}{\rwilRecallMCThreeVanilla$\pm$1}  & 
\multicolumn{1}{l|}{\rwilRecallMCTwoVanilla$\pm$1}      & 
\rwilRecallMCThreeVanilla$\pm$1                       \\ \cline{2-11} 

&\tool  & 
\multicolumn{1}{l|}{\bf \rwilRecallMOne$\pm$1}     & 
\multicolumn{1}{l|}{\bf \rwilRecallMTwo$\pm$1}  & 
\multicolumn{1}{l|}{\bf \rwilRecallMThree$\pm$1}  & 
\multicolumn{1}{l|}{ \rwilRecallCOne$\pm$1}  & 
\multicolumn{1}{l|}{ \rwilRecallCTwo$\pm$1}  & 
\multicolumn{1}{l|}{ \rwilRecallCThree$\pm$2}  & 
\multicolumn{1}{l|}{\bf \rwilRecallMCOne$\pm$1}  & 
\multicolumn{1}{l|}{\bf \rwilRecallMCTwo$\pm$2}      & 
{\bf \rwilRecallMCThree$\pm$2}                        \\ \hline

\end{tabular}
\label{table: rq2_real_instance}
\end{table*}

%% file: Tables/rq2_real_comparison_table.tex
\begin{table*}[]
\centering
\caption{Recall rates on \totalRealWorldModernStatic static methods moved by {OSS} developers in 2024. $Recall_M$ = identify the method, $Recall_C$ = identify the target class for a given method, $Recall_{MC}$= identify the method\&target class pair. }

% \todo{For \feTruth we report recall rates only for a subset of \feTruthProcessedRealWorld refactorings, for the remaining refactorings \feTruth did not recommend any refactorings within the 1-hour time-limit.}}
\begin{tabular}{|l|l|lll|lll||lll|}
\hline
\multicolumn{1}{|c|}{\multirow{2}{*}{Oracle Size}} & 
\multicolumn{1}{c|}{\multirow{2}{*}{Approach}} & 
\multicolumn{3}{c|}{$Recall_M$} & \multicolumn{3}{c|}{$Recall_{C}$} &  \multicolumn{3}{c|}{$Recall_{MC}$}                                                                \\ \cline{3-11} 
\multicolumn{1}{|c|}{} &\multicolumn{1}{c|}{}                          & \multicolumn{1}{c|}{@1} & \multicolumn{1}{c|}{@2} & \multicolumn{1}{c|}{@3} & \multicolumn{1}{c|}{@1} & \multicolumn{1}{c|}{@2} & \multicolumn{1}{c|}{@3} & \multicolumn{1}{c|}{@1} & \multicolumn{1}{c|}{@2} & \multicolumn{1}{c|}{@3} \\ \hline

% \JMove     &    \multicolumn{1}{l|}{-}     & 
% \multicolumn{1}{l|}{-}  & 
% \multicolumn{1}{l|}{-}  & 
% \multicolumn{1}{l|}{-}  & 
% \multicolumn{1}{l|}{-}  & 
% \multicolumn{1}{l|}{-}  & 
% \multicolumn{1}{l|}{-}  & 
% \multicolumn{1}{l|}{-}      & 
% -                        \\ \hline

\multirow{3}{*}{SmallClasses (40)} &
\feTruth    & 
\multicolumn{1}{l|}{7\%}     & 
\multicolumn{1}{l|}{15\%}  & 
\multicolumn{1}{l|}{15\%}  & 
\multicolumn{1}{l|}{14\%}  & 
\multicolumn{1}{l|}{14\%}  & 
\multicolumn{1}{l|}{14\%}  & 
\multicolumn{1}{l|}{1\%}  & 
\multicolumn{1}{l|}{2\%}      & 
2\%                  \\ \cline{2-11} 

&Vanilla-LLM & 
\multicolumn{1}{l|}{42\% $\pm$1}     & 
\multicolumn{1}{l|}{55\% $\pm$2}  & 
\multicolumn{1}{l|}{64\% $\pm$1}  & 
\multicolumn{1}{l|}{8\% $\pm$1}  & 
\multicolumn{1}{l|}{8\% $\pm$1}  & 
\multicolumn{1}{l|}{8\% $\pm$1}  & 
\multicolumn{1}{l|}{4\% $\pm$1}  & 
\multicolumn{1}{l|}{4\% $\pm$1}      & 
4\% $\pm$1                      \\ \cline{2-11} 

&\tool & 
\multicolumn{1}{l|}{\bf 54\% $\pm$1} & 
\multicolumn{1}{l|}{\bf 62\% $\pm$3}  & 
\multicolumn{1}{l|}{\bf 69\% $\pm$1}  & 
\multicolumn{1}{l|}{\bf 22\% $\pm$4}  & 
\multicolumn{1}{l|}{\bf 26\% $\pm$4}  & 
\multicolumn{1}{l|}{\bf 26\% $\pm$4}  & 
\multicolumn{1}{l|}{\bf 12\% $\pm$2}  & 
\multicolumn{1}{l|}{\bf 16\% $\pm$2}      & 
{\bf 18\% $\pm$2}                        \\ \hline \hline

\multirow{3}{*}{LargeClasses (62)} &
\feTruth   & 
\multicolumn{1}{l|}{6\%}     & 
\multicolumn{1}{l|}{11\%}  & 
\multicolumn{1}{l|}{15\%}  & 
\multicolumn{1}{l|}{6\%}  & 
\multicolumn{1}{l|}{6\%}  & 
\multicolumn{1}{l|}{6\%}  & 
\multicolumn{1}{l|}{0.4\%}  & 
\multicolumn{1}{l|}{1\%}      & 
1\%                  \\ \cline{2-11} 

&Vanilla-LLM & 
\multicolumn{1}{l|}{14\% $\pm$1}     & 
\multicolumn{1}{l|}{22\% $\pm$1}  & 
\multicolumn{1}{l|}{28\% $\pm$6}  & 
\multicolumn{1}{l|}{3\% $\pm$5}  & 
\multicolumn{1}{l|}{3\% $\pm$5}  & 
\multicolumn{1}{l|}{3\% $\pm$5}  & 
\multicolumn{1}{l|}{0.3\% $\pm$1}  & 
\multicolumn{1}{l|}{1\% $\pm$1}      & 
1\% $\pm$2                      \\ \cline{2-11} 

&\tool & 
\multicolumn{1}{l|}{\bf 17\% $\pm$4}     & 
\multicolumn{1}{l|}{\bf 28\% $\pm$4}  & 
\multicolumn{1}{l|}{\bf 32\% $\pm$4}  & 
\multicolumn{1}{l|}{\bf 43\% $\pm$5}  & 
\multicolumn{1}{l|}{\bf 43\% $\pm$5}  & 
\multicolumn{1}{l|}{\bf 47\% $\pm$5}  & 
\multicolumn{1}{l|}{\bf 7\% $\pm$1}  & 
\multicolumn{1}{l|}{\bf 12\% $\pm$1}      & 
{\bf 15\% $\pm$1}      \\ \hline

\end{tabular}
\label{table: rq2_real}
\end{table*}

%% file: Files/RQs/rq3_result.tex
\subsection{Runtime performance of \tool (RQ3)}
%We capture a profile of \tool's runtime performance through its pipeline.

% \subsubsection{Dataset} To evaluate the runtime performance of \tool, we used both the synthetic and real-world corpus employed in our earlier experiments. 

% This dataset, comprising a diverse range of synthetically created move method scenarios, provides a controlled environment for systematic performance testing across varying class sizes and complexities. 
% {\color{red}{Why not real-world dataset??}}

%{\bf Experimental Setup.} 

\subsubsection{Experimental Setup}

We used both the synthetic and real-world corpus employed in other RQs to measure the time taken for each tool to produce recommendations.
% \shorten{The execution time is an interval between the tool's triggering on a source class and the display of final refactoring suggestions to the user.} 
To understand what components of \tool take the most time, we also measured the amount of time it took to generate responses from the LLM, and the time it took to process suggestions.
To ensure real-world applicability, we conducted these measurements using the \tool plugin for IntelliJ IDEA, mirroring the actual usage scenario for developers. 
% \shorten{This approach allows us to account for any overhead introduced by the IDE integration, providing a more accurate representation of the tool's performance in practical settings.}
We conducted all experiments on a commodity laptop, an M1 MacBook Air with 16GB of RAM. 

%{\bf Results.} 

\subsubsection{Results}

Our empirical evaluation demonstrates that \tool achieves an average runtime of \mmAssistRuntimeAvg seconds for generating suggestions. The primary computational overhead stems from the LLM API interactions consuming approximately \llmRuntimeAvg seconds. 
% \todo{This is in line with XYZ guidelines for plugin development.}
% JMove takes about an hour
%On the contrary, JMove takes 48 minutes to complete its analysis ~\citep{TERRA201819JMove}.
In our experience with \JMove, on the larger projects in our real-world dataset, \JMove takes several hours (up to 24 hours) to complete running, thus we imposed the 1-hour cutoff time. {Similarly, \HMove also takes an {average of 80min} to execute on a single entry in our dataset --  it needs to be triggered on all possible $<$method, target class$>$ pairs ({avg. 145 pairs per class}).
% -- each taking several minutes on average.
} {In extreme cases where the host class was large (>10K LOC), \HMove took 4 whole days to execute on a single entry in our dataset.} Out of the box, \feTruth is also slow and can take 12+ hours to run on large projects. With the help of \feTruth authors, we were able to run it on a single class at a time -- this takes an average 6 minutes per class. {Thus, compared with the baselines, \tool is two, two, and one order(s) of magnitude faster than \JMove, \HMove, and \feTruth,~respectively. Thus, it is practical.}

% As a result, \mm spends \todo{p\%} of it's time waiting for LLM responses.

%% file: Files/RQs/rq4_result.tex
\subsection{Usefulness of \tool (RQ4)}
We designed a user study to assess the practical utility of \tool from a developer's perspective.

%{\bf Dataset.} 

\subsubsection{Dataset}

We made the deliberate choice to have participants use projects with which they were familiar. This decision was grounded in several key considerations. First, 
% by working with familiar codebases, developers can leverage their deep understanding of the project's architecture, design decisions, and evolution history. 
familiarity with their codebases enables them to make more informed judgments about the appropriateness and potential impact of the suggested refactorings. Second, using personal projects enhances the validity of our study, as it closely mimics real-world scenarios where developers refactor code they have either authored or maintained extensively. Third, this approach allows us to capture a diverse range of project types, sizes, and domains, potentially uncovering insights that might be missed in a more constrained, standardized dataset.

%{\bf Experimental Setup.}

\subsubsection{Experimental Setup}

% \todo{We invited to participate in our user study 43 graduate students (Master's and Ph.D.) taking the Software Refactoring course in Fall 2024 at \anonymizedUniversity University with Professor \anonymizedProfessor as the course instructor. The students were taught about the Feature Envy code smell with case studies from open-source projects a week before the user study started.
% The students were incentivized to participate in the user study with a 5\% bonus added to their overall course marks.} 
% Out of the initial pool of 43 students taking the course, 
% \commentb{We invited to participate in our user study 43 graduate students}
\numUserParticipants students (25 Master's and 5 Ph.D. students) volunteered to participate in our study. Based on demographic information provided by the participants, 
%50\% of them have professional industrial experience of +2 years.
% , 10 have less than 2 years of experience in the software industry, while 8 of them have no industrial experience. 
% 10 of them have industrial experience between 2-5 years, 10 of them have less than two years of industrial experience, 2 of them have more than five years of industrial experience, while 8 of them have no industrial experience. 
%Overall, 
73\% have  industrial experience.
All participants, with the exception of two, have experience with the Java programming language. 
% (13 students have 2 or more years of experience, and 15 students with less than two years of experience).
% (15 with less than two years of experience, 8 with 2-5 years of experience, and 5 with more than five years of experience).
Finally, the majority of participants (24 out of \numUserParticipants) have prior experience with refactoring.

We instructed the participants to use  \tool for a week and run it on at least ten different Java classes from~their projects. 
{The selection of these classes was left to the discretion of the participants, with the guidance to choose files they had either authored or were familiar with.} 
For each class they selected, \tool provided up to three \mm recommendations. We chose to present three recommendations to strike a balance between variety and practicality. 
% \todo{not to overwhelm users}. 

{The user study involved setting up our IntelliJ plugin, executing our tool on a class to get refactoring suggestions, rating the suggestions live inside the IDE, and lastly, providing feedback by filling out our survey.}
Afterward, they sent us the fine-grained telemetry data from the plugin usage. 
{For confidentiality reasons, we anonymize the data by stripping away any sensitive information about their code.
% , e.g., the names and source code of classes or methods that \tool presented to them. 
}
We collected usage statistics from each invocation of the plugin on each class. In particular, we collected this information: how the users rated each individual recommendation and whether they finally changed their code based on the recommendation.

Participants rated each recommendation on a 6-point Likert scale ranging from (1) Very unhelpful to (6) Very helpful. 
We chose this 6-point Likert scale to force a non-neutral stance, encouraging participants to lean towards either a positive or negative assessment. 
We asked the participants to rate the \tool's recommendations while they were fresh in their minds, right after they analyzed each recommendation.

After participants sent their usage telemetry, we asked them to fill out an anonymous survey asking about their experience using \tool. We asked participants to compare \tool's workflow against the IDE, and asked for open-ended feedback about their experience.

\subsubsection{Results}

\numUserParticipants participants applied \tool on \totalClassesUser classes. 
{\tool analyzed 1,143 host classes where developers did not have prior knowledge of refactoring opportunities. \tool recommended refactorings in 350 of them, with an average of 1.7 recommendations/class. \tool did not deem 793 typical classes to require refactoring, avoiding unnecessary developer effort.
}
% \commentb{\tool recommended 1.7 \mm on average.}
% Each user invoked the tool on \avgClassesUser classes on average. 
% They were presented with a total of \totalSuggestionsUser suggestions from \tool, with each user inspecting an average of \avgSuggestionsUser suggestions. 
% Of these suggestions, users rated \percentPositivePerSuggestion of the suggestions positively (`Somewhat Useful' or better). 
% As we provide multiple suggestions per class (upto 3 suggestions), we also wanted to understand if any one of the suggestions provided was rated positively. 
% To understand this, we kept the best-rated suggestion from each class and found out how many of them were positively rated. 
We found that, in \totalPositiveClassesUser classes
% (out of \totalClassesUser classes) 
the participants positively rated one of the recommendations (\percentPositivePerScreen of the time). 
% \percentPositivePerScreen of time. 
% \todo{Nikos: It is very unclear what is the difference between the percentages 74.5\% and 86.6\%. I cannot understand the difference.}
 Moreover, the users~accepted and applied a total of \totalAppliedUser refactoring recommendations on their code {(out of 354 total recommendations)}, i.e. \avgAppliedUser refactorings per user, on average. This shows that our tool is effective at generating useful recommendations that  developers, who are familiar with their code, accept.

The participants also provided feedback in free-form text. Of the \numUserParticipants participants, \percentPositiveRatingSurvey of them rated the plugin's experience highly, when comparing against the workflow in the IDE.
In praise of \tool, the participants said that \tool gave them a sense of control, allowing them to apply refactorings that they agreed with.

%% file: Files/threats.tex
\section{Discussion}
\label{sec:threats}

{{\bf Internal Validity: }}
Dataset bias poses a potential threat to the effectiveness of \tool. To mitigate this, we employ both a synthetic dataset (widely used by others), offering a controlled environment, \emph{and} a real-world dataset comprising refactorings performed by open-source developers. 

{\bf External Validity: } This concerns the generalization of our results. Because we rely on a specific LLM ({GPT-4o}), it may impact the broader applicability of our findings. We anticipate that advancements in LLM technology will improve overall performance, though this needs to be verified empirically. Second, \tool currently focuses on Java code. 
{Although our approach is conceptually language- and refactoring-agnostic, extending to additional refactoring types and languages requires adapting three key components: (1) static analysis for validating refactoring preconditions, (2) semantic analysis of code relationships, and (3) refactoring execution mechanics. Using protocols like the Language Server Protocol (LSP)~\citep{lspRepo} can simplify handling language-specific features, facilitating broader applicability.}
% \commentb{Although our approach is conceptually language-agnostic, with prompts not tied to Java-specific constructs, extending to other languages requires adapting three key components: (1) static analysis for validating refactoring preconditions, which can vary across statically and dynamically typed languages, (2) semantic analysis of code relationships, which must account for different module systems, and (3) refactoring execution mechanics, which need language-specific implementations. While these adaptations primarily involve engineering effort rather than conceptual redesign, each target language would require careful consideration of its unique type system, module organization, and refactoring constraints.} 
Future work will explore the effectiveness of our tool across various languages and refactorings. {While we utilize IntelliJ's refactoring engine for validation, we acknowledge it may contain latent bugs~\citep{refEngineBugs}. To mitigate this, \tool employs a sanity check to pre-filter invalid move methods. Therefore, the reliability of our approach depends not only on the correctness of IntelliJ's engine but also on the effectiveness of our pre-filtering stage. }

% This concerns the generalizability of our results. Because we rely on a specific LLM, it may impact the broader applicability of our findings. We anticipate that advancements in LLM technology will improve overall performance, though this needs to be verified empirically. Second, \tool currently focuses on Java code. Although our approach is conceptually language-agnostic, with prompts not tied to Java-specific constructs, we cannot definitively claim generalizability to other programming languages without further investigation. Future work will explore the effectiveness of \tool across diverse programming languages. % to establish its broader applicability in software engineering practices.

% \paragraph{\bf Verifiability:} 

{\textbf{Tool implementation.}}
% \todo{talk about how we could apply to other languages, and to other IDEs, or IDE independence through LSP}
\tool's implementation follows a modular architecture, separating language-specific concerns from the core refactoring workflow. Components such as the LLM service, embedding model, and IDE integration communicate via well-defined interfaces, facilitating extensibility and integration across various environments and languages. 
% \tool's architecture supports extensibility across programming languages and development environments through three fundamental design decisions. First, our core refactoring logic separates language-specific concerns from the refactoring workflow. While the current implementation uses IntelliJ's refactoring framework for Java, future versions could integrate with the Language Server Protocol (LSP)~\citep{lspRepo}. LSP is a standardized protocol that enables development tools to provide consistent programming language support across different editors and IDEs. Second, we designed a modular pipeline where components communicate through well-defined interfaces. The LLM service, embedding model, and IDE integration are independent modules that can be replaced without affecting the rest of the system. For instance, developers can swap different embedding models while retaining the same similarity computation logic or integrate different IDEs through their native plugin APIs. While our evaluation shows that general-purpose LLMs like GPT-4o, which also perform well at code understanding~\citep{arenaLeaderBoard}, our modular architecture allows easy integration of code-specific LLMs (e.g., StarCoder~\citep{li2023starcoder}, CodeLlama~\citep{roziere2023code}) through the same interface. This opens interesting research directions to evaluate whether models specifically trained on code repositories could offer better refactoring suggestions.  
% \todo{talk about how we deal with non-determinism}
% 
{To address the non-deterministic nature of LLMs, we experimented with various temperature settings and found the variability in the LLM's outputs to be consistently low, as evidenced by the small standard deviations reported in our evaluation.}
% Write about implication for IDE developers (moving to and from of generic classes)
For IDE developers, \tool shows the safe integration of AI-powered suggestions with existing IDEs.
% \tool is available~\citep{ReplicationPackage}. 

%% file: Files/related.tex
\section{Related Work}
\label{sec:related}
We organize the related work into: (i) research on \mm, and (ii) usage of LLMs for refactoring.

% tools that identify code smells and tools that automatically refactor code. In contrast, our tool not only identifies code that requires refactoring but also performs the refactoring automatically. While previous work includes few tools that can both identify and refactor code, these tools are often limited to specific types of refactorings. Our tool, however, is capable of identifying a wide variety of refactorings and automating the refactoring process.

% list of move method 

{\textbf{\mm refactoring.}}
% During a survey conducted by \citep{mens2004surveyofrefac}, it was found that
% the refactoring process involves several distinct activities: first, identifying areas of the software that require refactoring; second, determining which kinds of refactoring should be applied; third, ensuring that these refactorings preserve the existing behavior of the software; fourth, applying the chosen refactorings; fifth, assessing their impact on various quality characteristics, such as complexity, understandability, and maintainability, as well as on process metrics like productivity, cost, and effort; and finally, maintaining consistency between the refactored code and other software artifacts, including documentation, design documents, requirements specifications, and tests.
% \shorten{The refactoring process, as found in the survey conducted by \citep{mens2004surveyofrefac}, is a pipeline of steps that encompasses identifying areas for refactoring, selecting the appropriate refactoring type, ensuring existing software behavior is preserved, implementing changes, evaluating their impact on quality and process metrics, and maintaining consistency with related documentation and artifacts.}
{Many researchers focus on identifying and recommending \mm refactorings.} 
%\todo{While many research efforts have explored automating or partially automating some steps in this pipeline, we fully automate the entire pipeline for \mm}. 
\JMove~\citep{TERRA201819JMove}, JDeodorant~\citep{JDeodorant}, and MethodBook~\citep{Bavota:2014:MethodBook}
% detect methods located in incorrect classes and suggest moving them to more appropriate classes based on 
suggest refactorings based on software metrics derived from static analysis. {Additionally, \JMove introduced a widely-used synthetically created dataset of \mm refactorings.}
{\HMove~\citep{cui2024three}, a recently introduced tool, uses graph neural networks to classify a \mm suggestion as go/no-go. Then, \HMove only uses LLM as a judge to filter suggestions that don't meet certain preconditions}.
% , relying on several user-configured thresholds to detect and recommend move method refactorings. 
% \commentb{Works such as MethodBook~\citep{ Bavota:2014:MethodBook} compute project-level data structures to recommend \mm.} 
% \citep{2013WCRARecommendMove} recommends move method refactorings based on the set of static dependencies established by a particular method, while \citep{Bavota:2014:MethodBook} uses relational topic models to identify methods to move. \citep{2016SANERDomino} further expands on move method refactoring by identifying other methods that should be moved once a method is relocated by a developer. 
{Similarly,
% On the other hand
} 
\feTruth~\citep{feTruth}, RMove~\citep{RMove:2022}, and PathMove~\citep{ kurbatova2020recommendation}, utilize {DL techniques to identify \mm opportunities.}
% \commentb{MANTRA ~\citep{xu2025mantra}, a recent work, is also capable of suggesting \mm given the target method}. 
% \todo{MANTRA does not provide the \mm refactoring opportunities as in {\tool}}.
% both structural and semantic representations from code snippets and employs machine learning models to recommend move method refactoring opportunities. 

% \shorten{In contrast to our approach, a major difference with these tools is that they require users to define thresholds, which can be challenging for inexperienced users.}
% \commentb{Many of these techniques depend on user-defined thresholds to suggest refactorings, and only partially automate the refactoring process.}
% Most of these techniques recommend or partially automate refactoring activities but do not fully automate the whole refactoring pipeline~\citep{mens2004surveyofrefac}, unlike our approach.
Most importantly, \tool attacks the problem in a different direction. Previous tools compute whole project dependencies (which is computationally expensive and doesn't scale) and then produce a confidence score for each method in the project. 
Thus, they treat this as a \emph{classification}, not a \emph{recommendation} problem:
they produce {\overwhelmRecoCount} recommendations {on average} to move out a given class (many of which are unuseful), which puts tremendous analysis burden on programmers. In contrast, \tool offers at most 3 recommendations per class, {aligned} with how expert developers refactor code.

% \todo
{\textbf {Refactoring in the age of LLMs.}}
A recent systematic study~\citep{Hou2024TSE} analyzing 395 research papers demonstrates that LLMs are being employed to solve various software engineering tasks. 
%, such as software testing, code review, code summarization, and code generation, with their adoption significantly enhancing developer productivity~\citep{paradis2024much}.
%Among the earliest LLM-powered tools contributing to this progress is GitHub Copilot~\citep{copilot}, which provides real-time, context-aware code suggestions. Other notable tools include AlphaCode~\citep{AlphaCode} from DeepMind, CodeWhisperer~\citep{CodeWhisperer} from Amazon, Tabnine~\citep{Tabnine}, and Cursor~\citep{Cursor}, which focus primarily on code completion, editing, and generation.  
While code generation has been the predominant application, recently LLMs like ChatGPT have been applied to automate code refactoring~\citep{shirafuji2023refactoring,depalma2024exploring,AlOmar2024MSR,cui2024three,liu_exploring_2025} and detect code smells~\cite{detecting_code_marco}. 
Cui et al.~\citep{cui2024one} leverage intra-class dependency hypergraphs with LLMs to perform extract class refactoring, while iSMELL~\citep{isSMELL} uses LLMs to detect code smells and suggest corresponding refactorings. However, LLMs are prone to hallucinate, which can introduce incorrect or broken code, posing challenges for automated refactoring systems. 
Unlike other approaches, \tool addresses this limitation by validating and ranking LLM-generated outputs, ensuring that developers can safely execute refactoring recommendations. 

The prevalence of hallucinations in LLM-based refactoring is widely studied.
Pomian et al.~\citep{EM-Assist,PomianICSME2024} investigated hallucinations in \textsc{ExtractMethod} refactoring, while \citep{pycraft} analyzed hallucinations in Python code modifications. These studies consistently show that LLMs can hallucinate during refactoring tasks, substantiating our findings, where LLMs hallucinated in \hallucinationRate of the cases when suggesting \mm. This highlights the necessity of robust validation mechanisms, which are integral to our \tool, ensuring the reliability and safety of the suggestions generated by LLMs. 
{MANTRA~\citep{xu2025mantra} introduces a multi-agent approach with RAG to automate refactoring. However, MANTRA uses LLM to generate the refactored code directly, guiding this process with a RAG system that retrieves examples of correct refactorings from open-source projects to serve as few-shot prompts. In contrast, we leverage the LLM only to identify and propose refactoring candidates, while delegating the safe execution of the code transformation to the IDE. Our application of RAG also differs, as we use it to narrow down the search and retrieve a smaller number of potential target classes. }

%% file: Files/conclusion.tex
\section{Conclusion}
\label{sec:conclusion}

Despite years of research in \mm refactoring, progress has been incremental. The rise of LLMs has revitalized the field. Our approach and tool, {\tool}, significantly outperforms previous best-in-class tools and provides recommendations that better align with the practices of expert developers. When replicating refactorings from recent open-source projects, \tool achieves 4x higher recall while running 10x–100x faster. Additionally, in a one-week case study, \numUserParticipants experienced developers rated \percentPositivePerScreen of \tool~’s recommendations positively.

%Despite lots of active research in the area of recommending \mm refactorings, the progress over the years has been incremental and has stifled. The rise of LLMs and their applications to the field of refactoring has rejuvenated the field. Our approach and tool, \tool, significantly outperforms previous best-in-class tools and provides recommendations that better align with the practices of expert developers. When replicating a corpus of refactorings performed in 2024 by open source developers, \tool improves the recall over previous best-in-class tools by 4x, while running in 10x--100x less time. Moreover, our case study with \numUserParticipants experienced participants who used \tool to refactor their own code for one week shows they rated \percentPositivePerScreen of \tool recommendations  positively. 

The key to unleashing these breakthroughs is combining static and semantic analysis to (i) eliminate LLM hallucinations and (ii) focus its laser. \tool checks refactoring preconditions automatically which cuts down the LLM hallucinations. By leveraging semantic embedding into a RAG approach, \tool narrows down the context for the~LLM so that it can focus on a small number of high-quality prospects. This was instrumental in picking the right candidate from industrial large scale projects.  %containing more than 21000 candidate target classes. 
We hope that these techniques inspire others to solve other refactoring recommendation domains, e.g., splitting large classes or packages. 
% Thus, we make \tool and datasets available~\citep{ReplicationPackage}. 

\section{Data Availability}
\label{sec:data}

 To ensure the verifiability of our work and to provide a foundation on which others in the community can build upon, we have made our tool, \tool, publicly available in our replication package
% % publicly available as an open-source project
 ~\citep{ReplicationPackage}. This includes the complete source code, the IntelliJ plugin implementation, comprehensive documentation detailing the setup and usage instructions, a demo of \tool in action, the exact LLM prompts that we use, and all datasets we use in our evaluation.

% Additionally, we provide access to the datasets used in our evaluation, allowing other researchers to replicate our experiments and validate our findings.

\section*{Acknowledgements}

We are grateful for the constructive feedback from the members of the AI Agents team at JetBrains Research and the anonymous conference reviewers. This research was partially funded through the US National Science Foundation (NSF) grants CNS-1941898, CNS-2213763, 2512857, 2512858, the Industry-University Cooperative Research Center on Pervasive Personalized Intelligence (PPI), and a gift grant from NEC. Tien N. Nguyen was supported in part by the NSF grant CNS-2120386, and the National Security Agency (NSA) grant NCAE-C-002-2021.